\documentclass[12pt]{article}

\usepackage{amsfonts,latexsym,eucal,color,graphicx,epsfig,amssymb,amsmath}
\usepackage{gensymb}

\begin{document}

\vspace{2cm}
\begin{center}
{\Large {\bf Biological implications of possible unattainability of comprehensive,
molecular-resolution, real-time, volume imaging of the living cell}} \\
\vspace{2cm}
{\bf Hiroshi Okamoto}\\
\vspace{.5cm} 
Department of Electronics and Information Systems, Akita Prefectural University, Yurihonjo 015-0055, Japan\\
\vspace{.5cm}
okamoto@akita-pu.ac.jp\\
\vspace{1.5cm}
\end{center}
 
\begin{abstract}
Despite the impressive advances in biological imaging, no imaging
modality today generates, in a comprehensive manner, high-resolution
images of crowded molecules working deep inside the living cell in
real time. In this paper, instead of tackling this engineering problem
in the hope of solving it, we ask a converse question: What if such
imaging is \emph{fundamentally} impossible? We argue that certain
decoherence processes could be suppressed because the internal workings
of the cell are not being closely ``observed'' in the quantum mechanical
sense, as implied by the assumed impossibility of imaging. It is certainly
true that the ``wet and warm'' living cell should not exhibit quantum
behavior merely because of the lack of observation. Despite this,
we plow ahead to see what \emph{might} result from such absence of
the outward flow of information. We suggest that chaotic dynamics
in the cell could be quantum mechanically suppressed --- a known phenomenon
in quantum chaology, which is potentially resistant to various mechanisms
for the emergence of classicality. We also consider places where experimental
evidence for/against such a possibility may be sought.
\end{abstract}

\section{Introduction}

The central premise, or at least the central working hypothesis, of
modern biology states that all forms of life, including us, are molecular
machines. One might actually go further to assert that every form
of life is essentially a \emph{classical }molecular machine. A few
qualifications are needed for this stronger version. First, without
quantum physics we do not even have the stability of atoms and molecules,
or their chemical reactions, not to mention the non-infinite specific
heat of the vacuum. However, we may still be able to construct a \emph{classical
model} that describes interactions of molecules so that it successfully
captures essential dynamics of the biological machine; much like the
way workings of a transistor can reasonably be modeled with a few
characteristic curves and/or equivalent circuits, hiding the ``quantum-mechanical
layer'' \textemdash{} an abstraction layer in the computer engineering
sense \textemdash{} of silicon atoms and electron waves underneath.
Second, several emerging aspects of life being investigated under
the heading of ``quantum biology'' challenge the view that life
is basically classical.\footnote{This may not be the consensus view. There is at least one biology
textbook filled with quantum physics: W. Bealek, \emph{Biophysics:
Searching for Principles} (Princeton University Press, 2012).} These aspects include photosynthesis \cite{Fleming photosynthesis,photosynthesis theory}
and magnetic sensing by European robins \cite{european robins}. Although
these phenomena are intriguing, it is fair to say that these are either
a relatively short-time phenomenon where the associated energy quanta
$\hbar\omega$ far exceed the thermal energy $k_{B}T$ as in the former,
or a small-scale phenomenon occurring in --- presumably, because the
detail is not yet known --- space spanning at most a few molecules,
as in the latter. In other words, \emph{in retrospect}, these quantum
phenomena are found where they are expected, i.e. when the length
or time scale is small, or only modestly larger than the scale where
quantumness certainly is expected.

In this paper, we propose a new perspective to study the relevance
of quantum mechanics in biology, possibly extending the arena where
quantum physics plays a role. The main idea is the following. It is
extremely difficult to devise an experimental method that enables
comprehensive, molecular-resolution, real-time, volume-imaging (CMRV)
of the living cell. The word ``comprehensive'' is added to mean
that most molecules, as opposed to e.g. fluorescently labeled molecules
that are a small fraction of the whole, can be observed simultaneously.
True three dimensional (3D) imaging is meant by the word ``volume
imaging'', that has the ability to see \emph{internal} structures
and therefore goes beyond quasi-3D methods such as 3D topography enabled
by the atomic force microscope (AFM). The stated extreme difficulty
to devise a CMRV method is justified because otherwise somebody capable
would have invented it, considering the utility such a CMRV method
would offer to biology research. Turning this circumstance on its
head, if the difficulty is not of practical but of \emph{fundamental}
character, then little information about the internal states of the
cell is leaking to the outside environment because otherwise we would
be able to measure it in principle. If this reasoning is correct,
then the system is somewhat \emph{isolated}, that in turn means quantum
mechanically more coherent evolution is expected to a certain degree,
which we later attempt to estimate.

The idea, that there is some connection between a biological object
and quantum measurement on it, is by no means new: Already in 1932
Bohr discussed the connection noting that, in effect, a biological
object might ``hide its ultimate secrets'' because a high-resolution
measurement would destroy the object \cite{Niels Bohr}. Fast-forwarding
to modern times, Henderson noted that all high-resolution imaging
methods in biology, be it X-ray diffraction or electron microscopy,
are limited by radiation damage to the biological specimen \cite{Henderson}.
It is true that a variety of methods \textemdash{} X-ray crystallography,
cryoelectron microscopy, nuclear magnetic resonance (NMR) spectroscopy,
scanning probe microscopy and ``super-resolution'' optical microscopy
to name a few \textemdash{} have had remarkable successes in supporting
biology research. However, these are complementary and each of these
has a limited domain of applicability, for otherwise we would have
a different or shorter list of methods in the above. Furthermore,
as we will elaborate on later, each method has its own \emph{fundamental}
limits and whether CMRV will ever be possible is an open question.
In the present work, while some attempts are made to justify the fundamental
impossibility of CMRV, we will eventually \emph{assume} such impossibility
in order to see whether something interesting follows from the assumption.
Meanwhile, those who assume the opposite should be encouraged to develop
a CMRV method.

The idea, that quantum mechanics plays a significant role in biology
at a larger scale than what are already mentioned, has also long history.
Several authors that include prominent figures proposed since no later
than 1967 that the human brain might be quantum mechanically coherent,
although this certainly is far from the mainstream view \cite{Umezawa,ENM,Hameroff}.
While the brain is outside the scope of the present work, a work that
critically investigates the connection between the brain and quantum
physics is relevant \cite{Tegmark}. Briefly, this work concludes
that the brain is such a dissipative system that dissipation-induced
decoherence takes place far too quickly for any quantum coherence
to survive. The question of whether the whole biological \emph{cell},
as opposed to the whole brain, is quantum mechanically coherent has
also been asked \cite{Ogryzko}. Their work, which also is far from
the mainstream view, nonetheless contains at least one reasonable
point that the internal workings of the cell are ``observed'' by
the environment with possibly unconventional pointer states \cite{Zurek},
to which the wavefunction ``collapses''. Much as these studies are
inspiring, it is perhaps fair to say that these are controversial
at best, mainly because it is so unlikely to have quantum coherence
in such ``wet and warm'' physical systems. To see this, it is sufficient
to consider the amount of effort it takes to build any quantum-enhanced
instrument: That involves e.g. cooling, calibrating, vibration-damping,
removing external noise, characterising intrinsic noise, and so on
so forth. Nonetheless, we do not have a proof that quantum effects
--- of the sort we are discussing --- do not appear in wet-and-warm
systems, because we do not fully understand the nature of the quantum-classical
boundary.

The emergence of classicality is explained in several ways. First,
classical mechanics is in a sense the short-wavelength limit of quantum
mechanics and as such it emerges exactly the same way as ray optics
emerges from wave optics. In this view, classicality emerges whenever
the potential is smooth in the scale of the de Broglie wavelength.\footnote{Needless to say, quantum mechanics is not about classical waves. Hence
this point alone cannot capture the entirety of the quantum-classical
transition.} Second, things become classical when the action $S$ involved is
much larger than $2\pi\hbar$. Loosely speaking, when a characteristic
time scale of the system is $\tau$, with the associated characteristic
frequency scale $\omega\sim2\pi/\tau$, then classicality requires
$E\gg\hbar\omega$, where $E\sim S/\tau$ is a characteristic energy.
Quantization manifests itself when the particle energy such as $\hbar\omega$,
which tends to be larger for lighter particles, is larger than a characteristic
energy scale of the system. Such characteristic energy scale may be
the thermal energy $k_{B}T$ in a system embedded in an environment.
This explains why most quantum experiments demand cryogenic temperatures,
besides ones involving optical photons for example.\footnote{This is only a rule of thumb. See e.g. F. Galve, L. A. Pachon, and
D. Zueco, Bringing entanglement to the high temperature limit, Phys.
Rev. Lett. $\mathbf{105}$, 180501 (2010).} However, these conditions are still not complete: Microscopic quantum
uncertainty could always be amplified to a macroscopic quantum uncertainty
since quantum mechanics is linear, as exemplified by Schrodinger's
cat. The third way, in which classicality emerges, is decoherence
due to interaction with the environment \cite{Zurek Physics Today}.
In this view, \emph{we} as part of the environment get totally entangled
with Schrodinger's cat so that we do not notice the superposition.\footnote{If the reader is unsatisfied with this statement, he/she may put themselves
in the place of an external observer, so that ``we'' in the sentence
can be seen as only a bunch of atoms. Hence nothing more than the
usual ``shut-up-and-calculate interpretation'' is needed here. A
good explanation of quantum measurement that includes the observer
as a part of the system is found in a lecture note by Preskill. See:
J. Preskill, What is measurement?, Lecture Note, Ph 12b Quantum Physics,
California Institute of Technology, 28 Jan. (2010). Web address: http://www.theory.caltech.edu/\textasciitilde{}preskill/ph12b/Ph12b-measurement-28jan2010.pdf} The importance of \emph{isolation} of the system from the environment
is emphasized here. For example, setting aside the subtlety due to
the highly mixed nature of the states, NMR quantum computing works
because of isolation, in spite of the condition $k_{B}T\gg\hbar\omega$.
One way the amplification of minute quantum uncertainty happens drastically
is when the system's classical dynamics is chaotic; because the trajectory
depends on the initial condition in an exponentially sensitive way.
For example, amplification of quantum uncertainty to the macroscopic
level in a surprisingly short time has been discussed with respect
to very macroscopic objects including the Saturn's satellite Hyperion
\cite{Zurek Physica Scripta}. It is only the interaction with the
environment, partly in the form of impinging photons in this case,
that keeps Hyperion classical. In the present work, we look into the
possibility that the biological cell is classically chaotic and is
not that well-observed by the environment as Hyperion is. Specifically,
we consider possible delocalization of biological molecules in the
cell, however provocative this possibility may seem.

This paper is organized as follows. A variety of biological imaging
methods, established ones and also proposed ones, are discussed in
Sec. 2 to show that the unattainability of CMRV is a serious possibility.
In Sec. 3 we first estimate how much ``information'' leaks out to
the environment from the living cell. We then consider how these findings
fit with what we know in quantum chaology including, in particular,
quantum suppression of classical chaos. We consider potential observational
or experimental consequences in case such quantum suppression of chaos
is indeed at work. Section 4 concludes the paper. Throughout the paper
we will use the SI system of units. Symbols $c$, $\varepsilon_{0}$,
$\hbar$ and $k_{B}$ respectively represent the speed of light, the
vacuum permittivity, the reduced Planck's constant and the Boltzmann
constant, of which some have appeared already. Some symbols such as
$\lambda$ stand for multiple things (e.g. wavelength and Lyapunov
exponent) in this paper to avoid cluttered presentation. The meanings
of such symbols should be clear from the context.

\section{Biological imaging methods and their limits}

In this section, we intend to show that \emph{un}attainability of
CMRV of the cell is a serious possibility. Beautiful results from
biology research can be at the same time misleading because they may
give an impression that any biological measurement is possible.\footnote{At the optimistic end of the spectrum, it has been mentioned, for
example, that an army of ``nanobots'' could go into someone's brain
to gather data. It should also be noted that such a possibility has
not been ruled out.} However, on a closer look one finds that each reported measurement
was preceded by setting up particularly favorable conditions generally
incompatible with other measurement methods, and furthermore each
measurement method is associated with various fundamental limits.
Hence it is entirely possible that CMRV is impossible.\footnote{It is unlikely that we actually can give a proof that CMRV is impossible
even if it is. Indeed, not all impossibility should necessarily have
a short and clever proof, like the one showing the impossibility to
solve the halting problem in computability theory. However, a lengthy
proof of the impossibility of CMRV could actually \emph{exist}, which,
however, may be beyond our reach in any practical sense. For example,
given a reasonable laboratory size, say a cube with $10\mathrm{km}$
edge length, there are only \emph{finite} ways to place atoms in the
laboratory if we digitize the spatial coordinates with a sufficiently
fine resolution (say, 10 pm). Almost all atomic configurations are
hopelessly messy and unstable, but some configurations contain scientific
equipment. However, if \emph{all} configurations do not offer CMRV,
then CMRV is impossible at least in a laboratory with a reasonable
size. (For the sake of completeness, we note that even if such a CMRV
configuration exists, it may not be \emph{constructible} from a realistic
initial state of the laboratory. Another minor point is that the size
of the cube is set to be so large because we have the X-ray free electron
laser in our mind.) Of course, if there is a \emph{principled reason}
why CMRV is impossible, which certainly is a possibility, then the
proof should be shorter than the brute-force kind mentioned above.
For this viewpoint, see: D. Deutsch,\emph{ The Beginning of Infinity:
Explanations That Transform the World} (Goodreads, 2011).}

We examine several biological measurement methods in the following.
A particular emphasis is on methods under development. While we also
try to present a broad perspective, we do not intend to give a full
review of established methods that are already widely used.

\subsection{X-ray}

X-ray methodologies used in biology are diverse. Some use soft X-rays,
especially in the so-called ``water window'' at the wavelength of
$2.34\mathrm{nm}\sim4.4\mathrm{nm}$, and others use hard X-rays.
Some use optical elements such as lens and mirror, while others including
crystal diffraction are lensless. While there have been much progress
on all these fronts \cite{X-ray imaging}, it remains true that the
X-ray damages biological specimen generally more than electrons do
for a given amount of structural information retrieved, mainly by
photo-ionization \cite{Henderson}. X-ray photons simply are so energetic.
Hence the specimen is destroyed before a sufficient number of X-ray
photons are collected to form an image. Whereas atomic-resolution
structures of biological molecules are routinely obtained in X-ray
crystallography because a large number of copies of the same molecules
are involved, when it comes to single, unique objects such as a structure
in a cell, X-ray methods are generally no better than the electron-based
counterpart in terms of radiation damage \cite{Henderson}.

The advent of the method called \emph{flash X-ray imaging} \cite{Flash X-ray original proposal,Neutze Flash X-ray proposal}
makes an interesting twist to the above picture. The X-ray free electron
laser (XFEL) invented long ago \cite{Madey XFEL proposal} has come
to fruition after many years of effort and is capable of producing
intense and short ($\sim\mathrm{fs}$) X-ray pulses. The idea is that
the inertia of atomic mass is no longer negligible at the femto-second
time scale and hence the atomic nuclei stay where they are during
the X-ray irradiation; despite the ionization of atoms due to the
intense X-ray. Thus, a diffraction image can be recorded ``before''
the specimen is destroyed. Perhaps surprisingly to the uninitiated,
phase information is generally in the diffraction pattern to allow
for reconstruction \cite{Sayre oversampling,Miao oversampling}. Nevertheless,
$10^{5}\sim10^{6}$ identical copies of the molecule (or molecular
complex) are still needed to get a 3D ``near-atomic resolution''
structure with currently foreseeable technology because each image
is essentially 2D \cite{Miao review}. However, 3D imaging using a
single copy of the specimen must be possible to begin to realize CMRV.
(Those who think that CMRV is obviously impossible because the X-ray
destroys the specimen, will find an argument based on \emph{reconstruction}
of the specimen below.) One way might be the use of tomography with
multiple and almost simultaneous X-ray pulses from different angles.
Another possible way is the use of ``ankylography'', whose feasibility
is under active research \cite{Sparse ankylography}.

Assuming that 3D image acquisition is possible in flash X-ray imaging,
we examine whether this imaging method would make CMRV possible. First,
we get one glaring point out of the way: X-ray flash imaging seems
to be fundamentally destructive for real-time imaging because the
specimen would be evaporated after filming only a single frame of
the movie. However, there is a rather strange possibility: We could,
in principle, \emph{reconstruct} the specimen each time, based on
the recorded full structural information, possibly using a molecular-precision
synthetic biology method \cite{Okamoto synthetic biology}. Despite
the progress in synthetic biology, however, no method today is capable
of building things out of molecular parts in a well-controlled way.
However, if we assume the realizability of such a method, one could
in principle produce a movie of a biological process with a time step
$\Delta t$ as follows: We first bring a frozen specimen to room temperature
for the period $\Delta t$, which is followed by rapid vitrification
of the specimen by cooling it down to a cryogenic temperature again
(see discussions on cryoelectron microscopy below). X-ray flash imaging
is then performed to generate structural data of the specimen. The
data in turn is used to reconstruct, by a cryogenic process, the next
specimen to repeat the process. Evidently, this is speculative at
best and the minimum possible interval $\Delta t$ perhaps is not
short enough for some important scientific questions. Since this notion
will appear repeatedly in this paper, we call this method reconstruction-based
cryogenic stop-motion photography (RCSP).

Second, the larger the number of voxels is, the more intense the needed
X-ray pulse is, in order to have a sufficient number of photons to
acquire enough information. However, the oscillating electric field
of the X-ray field cannot be arbitrarily large for the following reason.
To observe atomic details with diffraction, we need X-ray with the
wavelength of about the Bohr radius $a_{0}=4\pi\varepsilon_{0}\hbar^{2}/m_{e}e^{2}$,
where $m_{e}$ is the electron mass. According to the standard approximation,
an electron bound to an atomic nucleus is regarded free because the
X-ray frequency is much larger than the ``orbital frequency'' of
the bound electron. Hence classically we have 
\begin{equation}
m_{e}\ddot{x}=eE\cos\left(\omega t\right),\label{eq:X-ray shaking of electron}
\end{equation}
where $\omega=2\pi c/\lambda=2\pi c/a_{0}=cm_{e}e^{2}/2\varepsilon_{0}\hbar^{2}$
is the angular frequency of the X-ray and $E$ is the associated electric
field amplitude. Plugging $x=-x_{0}\cos\left(\omega t\right)$ in
Eq.(\ref{eq:X-ray shaking of electron}), we obtain $x_{0}=eE/m_{e}\omega^{2}$,
which is the motional amplitude of the electron being ``shaken''
by the X-ray field. A problem begins to emerge when the X-ray field
is so strong that $x_{0}$ approaches the Bohr radius $a_{0}$, because
then the electron approaches the neighboring atoms. Up to a numerical
constant of the order of $1$, the electric field at which $x_{0}\sim a_{0}$
is $E_{\mathrm{shake}}=m_{e}^{2}c^{2}e/4\pi\varepsilon_{0}\hbar^{2}\cong10^{16}\mathrm{V/m}$.
The flow of energy $J$ due to such an X-ray pulse is 
\begin{equation}
J=\varepsilon_{0}E_{\mathrm{shake}}^{2}c\sim10^{29}\mathrm{W/m^{2}}.\label{eq:X-ray shaking power flux}
\end{equation}
On the other hand, Fig. 2 of Ref.\cite{Miao review} shows that the
Linac Coherent Light Source (LCLS) has the X-ray brilliance of $\sim10^{24}\mathrm{(photon/s/0.1\%bw/mm^{2}/mrad^{2})}$.
A decent diffraction experiment should require $0.01\%$ bandwidth
monochromaticity and $0.1\mathrm{mrad}$ beam divergence because both
the monochromaticity and the beam divergence need to be roughly the
wavelength divided by the specimen size, where the specimen size is
$\sim1\mu\mathrm{m}$ in our case of single biological cells. Note
that the coherence requirements become increasingly demanding for
larger specimens. Hence we have $J\sim10^{18}\mathrm{W/m^{2}}$ since
each X-ray photon has energy of $\sim12\mathrm{keV}\sim10^{-15}\mathrm{J}$.
On the other hand, Ref.\cite{Miao review} states towards the end
that one needs $10^{5}\sim10^{6}$ specimens and $100$-fold increase
of the X-ray peak power to enable atomic resolution 3D imaging. Taking
this at the face value, this would mean up to 8 orders of magnitude
increase of the X-ray power might be needed if we want to obtain an
one-shot diffraction image with a single specimen, resulting in a
figure $J\sim10^{26}\mathrm{W/m^{2}}$. Comparing with Eq.(\ref{eq:X-ray shaking power flux}),
we have leeway of only a few orders of magnitude. This crude calculation
suggests that at least more careful evaluation is desired for the
flash X-ray method if it aspires to enable CMRV. Even assuming this
is not a problem, the electron-positron pair production begins at
the Schwinger limit $E_{S}=E_{\mathrm{shake}}/\alpha$, where $\alpha\cong\left(137\right)^{-1}$
is the fine structure constant \cite{Schwinger limit}. This suggests
a possibly general trend that higher power brings in new physics that
has to be dealt with.

\subsection{Electron microscopy}

Despite the impressive progress in biological electron microscopy
(EM), the resolution-limiting factor of biological EM remains to be
\emph{radiation damage} to the specimen by the probe electrons.\footnote{A historical note: Whereas R. P. Feynman expected electron microscopy
to be an important tool in biology, saying ``It is very easy to answer
many of these fundamental biological questions; you just look at the
thing!''; D. Gabor responded to L. Szilard's suggestion of biological
electron microscopy with dismissive ``What is the use of it? Everything
under the electron beam would burn to a cinder!'' As noted in the
main text, the tension between these two views persists today. These
quotes are taken respectively from: R. P. Feynman, There's plenty
of room at the bottom, presented at Annual Meeting of APS, Pasadena
(1959). Transcript: J. Microelectromech. Syst. $\mathbf{1}$, 60 (1992);
R. M. Glaeser, Cryo-electron microscopy of biological nanostructures,
Phys. Today $\mathbf{61}$, 48 (2008).} Hence the situation is qualitatively similar to the case of X-ray.
However, flash imaging with electrons seems somewhat contrived at
least at the fundamental level since the electrons are fermions and
moreover they electrostatically repel each other. In this subsection,
in spite of the growing relevance of e.g. serial block-face scanning
EM (SEM) in the brain ``connectmics'' research \cite{Serial block face SEM},
we focus on transmission EM (TEM) that is relevant in molecular and/or
atomic scale imaging.

The art of biological TEM is multifaceted. The traditional methods
use protocols such as plastic embedding and heavy metal staining.
The reason for the latter is that the Coulomb force from the atomic
nuclei is the main scattering mechanism in TEM. These protocols work
excellently and led to many important discoveries at the scale of
organelles, but they introduce artefacts at the molecular scale. Thus,
cryoelectron microscopy (cryoEM) \cite{Glaeser Taylor cryoEM,Dubochet cryoEM}
was introduced in order to enable molecular or atomic scale resolution
in biological TEM. CryoEM uses ``frozen'' specimens at cryogenic
temperatures (typically the liquid nitrogen temperature) and hence
the specimen can be placed in a vacuum in its ``original'' form.
In order to prevent the growing ice crystals from destroying fine
biological structures, the method uses liquid ethane, to which the
thin specimen is rapidly dropped. The reason is that liquid ethane
tends to have less bubbles than liquid nitrogen does and hence allows
for a better thermal contact. Hence the specimen is rapidly vitrified
and the water remains in the glassy state. It is widely believed that
the specimen prepared as mentioned represents a biologically relevant,
natural structure. However, the contrast is very weak in cryoEM because
typical biological specimens contain mostly light elements. Moreover,
one can use only a limited dose of electrons of at most $\sim10{}^{3}e/\mathrm{nm}^{2}$
because of \emph{radiation damage} to the specimen caused by the very
electrons used for imaging. Hence, the resolution is limited by shot
noise, which is a fundamental consequence of the electron being a
particle.

Before proceeding, we comment on TEM in the broader context. TEM has
been used also in materials science since the beginning, where the
resolution had long been limited by lens aberration. The aberration
has finally been corrected in the 90's \cite{Haider Rose,Krivanek aberration correction},
fifty years after Scherzer showed possible solutions \cite{Scherzer aberration correction}.
Atomic resolution images are now routinely obtained for beam-\emph{in}sensitive
inorganic specimens. Furthermore, a study using aberration-corrected
TEM presents beautiful pictures of organic molecules in the carbon
nanotube \cite{Koshino Science}. Low-voltage, aberration-corrected
electron microscopes have been developed around the globe, enabling
imaging of beam-sensitive specimens \cite{JEOL LV aberration corrected TEM,nion LV aberration corrected TEM,Kaiser LV AC TEM}.
However, few things should be kept in mind. First, aromatic molecules
or ones with a similar structure, such as the nanotube, are very stable
and highly resistant against radiation damage; while most biological
molecules are not. Second, the specimen used in Ref.\cite{Koshino Science}
is very thin, whereas molecules of biological interest can be as large
as $30\mathrm{nm}$ and the vitrified specimen must be thick accordingly.
However, a thicker specimen requires higher beam energy in TEM, which
could produce knock-on damage. More importantly, the secondary electrons
generated by single inelastic scattering events produce more damage
in thick specimens. In short, beautiful pictures of organic molecules
in nanotubes do not mean that the same quality of images can be obtained
in other situations such as cryoEM of biologically interesting molecules.

At least two ways to bypass the radiation damage problem have been
successfully developed and these are established methodologies in
structural biology today \cite{Glaeser cryoEM book}. Both of these
are based on averaging to improve the signal-to-noise (S/N) ratio
using multiple copies of the same molecule. The first method is electron
crystallography that often use 2D crystals, which are not easy to
prepare. The second method is called ``single-particle analysis'',
where many identical molecules are prepared in vitreous ice \emph{without}
crystallizing them. Although each molecule has a random orientation,
approximate orientations of the molecules may be determined from the
noisy raw data, thus allowing us to average images of similarly-oriented
molecules, resulting in a better S/N of the estimated structure. This
in turn allows us to obtain a better estimation of molecular orientations
by fitting the estimated structure to the raw data. Iterating this,
one can increasingly refine the computed structure; although obviously
there are pitfalls associated with this clever method that generates
useful data from very noisy raw data. The uninitiated reader should
note that intricate structures obtained with cryoEM found in the scientific
literature are likely to be the result of such averaging and data
processing. In such cases, these structures do \emph{not} represent
the raw resolving power of cryoEM. The resolution of raw data remained
to be $5\sim10\mathrm{nm}$ because of radiation damage at least until
recently, when improved electron detectors began to be widely used
\cite{Glaeser nature methods}.

Ideas for potential future improvements to cryoEM are in the pipeline.
The so-called ``interaction-free measurement'', which is capable
of detecting an photon-absorbing object without actually getting photons
absorbed, was first proposed long ago \cite{Elitzur Vaidman Bomb}.
Subsequently, it was significantly refined so that the failure probability
can be brought down arbitrarily close to zero \cite{Kwiat-Kasevich},
and finally its use in biological EM was proposed \cite{Putnam-Yanik}.
While interaction-free measurement is inspiring, its use is limited
when it comes to distinguishing two semi-transparent objects, as opposed
to distinguishing the vacuum and a semi-transparent object \cite{Semitransparent IFM}.
However, this aspect is important because one would like to distinguish,
e.g. protein and water in cryoEM. Another proposed way to improve
cryoEM is the use of superconducting qubits to achieve measurement
at the Heisenberg limit, rather than at the the shot-noise limit \cite{Okamoto possible,Okamoto-Nagatani}.
This latter scheme may improve the resolution of cryoEM but it will
likely fall short of achieving the atomic resolution.\footnote{A somewhat outdated ``review'' of this particular proposal is presented
in Sec. I of: H. Okamoto, Measurement errors in entanglement-assisted
electron microscopy, Phys. Rev. A $\mathbf{89}$, 063828 (2014). On
a personal note, this author began suspecting the impossibility of
CMRV after spending a decade to learn that improving cryoEM in a fundamental
manner, even \emph{in theory}, is rather hard; although an appreciable
improvement turned out to be possible at least in theory.} The simple reason, after all, is that imaging methods using electrons
are fundamentally destructive. A possible way to get around this difficulty
would be the use of RCSP. Again, this possibility is speculative at
best.

\subsection{NMR}

NMR occupies a unique position in biological imaging because it does
not belong to the usual category where the incident wave scattered
by a specimen is measured. The frequency $\nu$ of the electromagnetic
wave is at most on the order of $1\mathrm{GHz}$ and hence the photon
energy $h\nu$ is at most $\sim10\mu\mathrm{eV}$, which is far smaller
than energy scales associated with an X-ray or an electron beam, or
even $k_{B}T$ at room temperature. Consequently, specimen damage
is minimal, if not zero, in NMR and this represents perhaps the most
attractive aspect of NMR. The downside of this is the weak signal
to be detected, which typically translates to a slow data acquisition
rate.

We first broadly review relevant aspects of NMR. As is well known,
volume imaging with NMR is routinely done today with clinical magnetic
resonance imaging (MRI), where the spatial resolution is typically
$\sim1\mathrm{mm}$. Using dedicated instruments, the MRI resolution
has been pushed down to $\sim1\mu\mathrm{m}$ for small specimens
\cite{Conventional MRI microscopy}. We also note that NMR has widely
been used to determine protein structures, using a solution containing
a massive number of the same molecule of interest. A number of techniques
are associated with such methods and reviewing NMR spectroscopy in
a decent way is firmly beyond the scope of this paper. Hence, in what
follows, we focus on methods that seem relevant to CMRV.

One method that aspires to enable CMRV is magnetic resonance force
microscopy (MRFM) \cite{Sidles MRFM proposal,MRFM early demonstration}.
The method has been improving ever since its introduction, resulting
in a relatively recent demonstration of biological imaging with resolution
$<10\mathrm{nm}$ \cite{Degen virus MRFM}. A conceptually similar
yet distinct method utilizing the nitrogen-vacancy (NV) center in
diamond has also been introduced recently \cite{NV-center first proposal and demo}.
This latter technology, NV magnetometry (NVM), has several distinct
usages but here we focus on the one relevant to CMRV, which is nano-scale
MRI. We consider whether MRFM/NVM represent a possible path to CMRV.
Since future prospects of the methods are nicely summarized in Ref.\cite{MRFM review}
at least in the case of MRFM, here we point out only few potential
obstacles. First, MRFM/NVM is a \emph{scanning} NMR method and hence,
combined with the fact that the weak signal requires integration for
a long time, it is slow. This makes real-time imaging difficult, unless
RCSP is employed. In any event, high-resolution MRFM/NVM measurements
are likely to eventually require cryogenic temperatures to reduce
thermal noise and to prevent diffusion of biological molecules. Second,
MRFM/NVM employs a magnetic field gradient $\partial B/\partial z$
as in the conventional MRI. The magnitude of the gradient $\partial B/\partial z$
determines the spatial resolution $\Delta z$, in such a way that
the field difference 
\begin{equation}
\Delta B=\frac{\partial B}{\partial z}\Delta z\label{eq:magnetic field gradient}
\end{equation}
corresponds to the attainable frequency resolution $\Delta f$ through
the gyromagnetic ratio. There are few possibilities that fundamentally
limit the frequency resolution $\Delta f$. First, there is the frequency
broadening due to dissipation, that lowers the $Q$ value of the oscillator
in MRFM, or in the case of NVM that determines the phase coherence
time of the NV center. It is hard to know if such dissipation is intrinsic
or extrinsic. Second, there is no motional narrowing in solid state
NMR, unlike in bulk liquid NMR. Consequently, almost inevitably we
have inhomogeneous broadening $\Delta f$ (besides intrinsic broadening
due to dissipation) of the resonance peak \cite{Solid state NMR}.
This broadening translates to a largely uncontrollable/unknowable
frequency shift of individual nuclear spins (typically that of hydrogen)
placed in the inhomogeneous environment, in localized measurements
such as ones performed in MRFM/NVM. The spatial resolution limit is
reached when such a shift $\Delta f$ matches $\Delta B$ in Eq.(\ref{eq:magnetic field gradient}).
On the other hand, the magnetic field gradient is typically produced
by a tiny, but strong, magnet placed nearby the biological specimen.
Naturally, the field gradient $\partial B/\partial z$ decays quickly
as the distance from the magnet increases. This in turn means that
we lose resolution fairly quickly as the depth from the specimen surface
increases. Hence volume imaging is hard with MRFM/NVM. Consequently,
even if MRFM/NVM will eventually be capable of atomic resolution imaging,
that will likely be confined at near the surface of the specimen.
While there are theoretical proposals \cite{NV center improvement proposal}
to address this problem, at present no proposal claims to achieve
atomic resolution imaging for the specimen of the size of the whole
cell, i.e. $>1\mu\mathrm{m}$.

\subsection{Scanning probe microscopy}

Among many variants in the category of scanning probe microscopy (SPM),
we focus on AFM here, because of the obvious relevance. In AFM, the
size of the force can be in the $\mathrm{pN}$ range and the oscillation
amplitude of the cantilever can be $\sim0.1\mathrm{nm}$. Hence, the
energy scale is $\sim10^{-22}\mathrm{J}$, which is less than $k_{B}T$
at room temperature. The energy is dissipated at the spatial scale
of atoms and at the time scale of mechanical oscillations (phonons),
similar to those of thermal molecular motions. These facts seem to
make AFM nondestructive. A remarkable image of an organic molecule
was obtained in 2009 \cite{Meyer organic molecule AFM}, despite under
a condition that does not seem to be extendable to general biological
imaging. In the following year, another remarkable result visualizing
a walking molecule in real time with a high-speed AFM was reported
\cite{Realtime AFM}. This has \emph{almost} realized CMRV, except
the ``volume imaging'' part. Hence the relevant question is whether
AFM would enable volume imaging. \emph{In theory}, one might remove
a thin layer after another as in serial block-face SEM, perhaps with
an ion beam of helium or argon, to study the 3D structure at a cryogenic
temperature. However, this will come at the price of losing true real-time
imaging at room temperature, even if comprehensive molecular resolution
imaging on etched surfaces is possible. Although the latter imaging
incorporating RCSP may seem possible \emph{in principle}, effects
such as charging of the surface may well make it impossible, as perhaps
most people experienced with imaging a thick piece of insulator with
AFM could attest.\footnote{What initially appears to be a mundane, practical kind of difficulty
may well be fundamental impossibility if it persists despite our best
effort. See also the argument in Footnote 6.} Another limiting factor in AFM is dissipation, which appears to be
a ubiquitous obstacle in advanced measurement in general, that broadens
the resonance peak of the cantilever. In summary, while AFM is likely
to be nondestructive, it is fundamentally surface specific. While
further research is desirable, at present no reliable path along the
use of AFM can be identified towards realizing CMRV.

\subsection{Optical microscopy}

Optical microscopy is a vast subject and we discuss only a few topics
here. We first discuss far-field methods and then near-field methods.
We also comment on the recently-introduced expansion microscopy.

Super-resolution microscopy methods are among those that are designed
to break Abbe's diffraction barrier $d=\lambda/2n\sin\alpha$ \cite{Super-resolution review}.
Most super-resolution methods rely on the use of fluorophores. One
group of super-resolution methods exploits non-linear properties of
the fluorophore. For example, in stimulated emission depletion (STED)
microscopy, the excited fluorophores are partially de-excited before
fluorescence takes place, leaving a small group of still excited fluorophores
with a size much less than the Abbe's limit. The de-excitation process
can be made non-linear so that the ``edge'' of the de-excited group
of fluorophores is sharp. The second group of method is based on the
observation that one can determine the \emph{center} of a point-like
optical image from a fluorophore to a precision much better than Abbe's
limit by fitting the point spread function. This, by itself, does
not allow for identifying multiple optical sources within the diffraction
limit, but if there is a way to identify each optical source \textemdash{}
for example by color, but the ``signature'' can be more subtle \textemdash{}
then one can determine the position of each source to a precision
that far exceeds Abbe's limit. Nonetheless, obstacles remain towards
CMRV. First and most obviously, only fluorophore-tagged molecules
can be observed and hence the ``comprehensive'' part of CMRV is
not satisfied. Even if all different kind of molecules were tagged
with different fluorophores, aside from the slowing down of the data
acquisition rate, questions would remain as to whether such a system
is natural. Second, at present the cell to be imaged are typically
fixed chemically, and fluorophores are overexpressed \cite{Betzig nobel lecture}.
This again raises a question as to whether the system is natural enough.
Third, optical bleaching of fluorophores defines the allowable photon
dose. This is a resolution limiting mechanism similar to the liming
factor in flash X-ray imaging and cryoEM. Moreover, the typical optical
intensity used in super-resolution imaging can be quite high for the
biological specimen itself. We note, in passing, that super-resolution
microscopy is not the only way to improve far-field optical microscopy.
Improvements exploiting quantum effects have been reported \cite{Taylor quantum optical microscope,Ono entanglement enhanced microscope}
but assessing these methods are beyond the scope of this paper.

Next, we discuss near-field methods. A representative example of these
methods is the scanning near-field optical microscope (SNOM) \cite{Ash-Nicholls SNOM}.
This is a scanning method and intrinsically slow. Moreover, NSOM captures
evanescent field for high resolution information and hence it is surface
specific. These two points, especially the latter, seem to render
SNOM hopeless as a candidate for CMRV. On the other hand, the advent
of the ``perfect lens'' \cite{Pendry perfect lens} presents an
interesting opportunity where \emph{all} evanescent field, including
ones emanating from deep inside the specimen, could be collected to
form an image. Nevertheless, the evanescent field does decay exponentially.
For ensuing discussions, the reader is referred to Ref.\cite{Near sighted perfect lens}.

Finally, we consider expansion microscopy introduced recently \cite{Expansion microscopy}.
This notable method makes the specimen \emph{actually} bigger, instead
of making the specimen \emph{look} bigger. This method is fundamentally
destructive, which is analogous to flash X-ray imaging or cryoEM.
Hence, even if it proves to be a comprehensive molecular-resolution
volume-imaging method in the future, it must incorporate RCSP to enable
CMRV. To reiterate, RCSP is at best a speculative possibility.

\medskip{}
 Two final remarks are in order. First, Henderson notes \cite{Henderson}
that neutrons damage the specimen little for a given elastic scattering
signal compared to the X-ray photons and electrons. He further notes
that the problem is of a practical character, which is that the brightness
of available neutron sources are many orders of magnitude smaller
than what would be needed and hence atomic resolution neutron microscopy
is ``impossibly far off''. It remains to be seen if this difficulty
is indeed of a practical character, or this is a fundamental problem
masquerading to be a practical one (see Footnote 9). Additionally,
deuterization needed for this imaging modality affects the functioning
of the cell. It is known that cells malfunction with heavy water \cite{deuterization of the cell}
and mammals die when they are fed heavy water extensively \cite{Heavy water poisoning}.
Second, discussions regarding acoustic microscopy are presented in
the next section.

To summarize, X-ray and electron-beam methods are fundamentally destructive
and they would require RCSP, a highly speculative possibility at present,
to enable CMRV. On the other hand, NMR-based methods, while non-destructive,
appears too slow to enable real-time imaging unless RCSP is employed.
Moreover, these methods appear to be fairly surface specific for reasons
discussed above. In scanning probe microscopy such as AFM, true-3D
imaging appears out of reach although interesting real-time results
have been obtained for molecules on the surface. Finally, in optical
microscopy, fluorophores are required in a condition that may be far
from the physiological condition, casting doubt as to whether optical
microscopy will ever enable CMRV. Thus, we conclude that the present
state of the art of microscopy does \emph{not} reject the hypothesis
that CMRV is fundamentally impossible.

\section{Quantum physical implications of the unobservability}

In this section, we further investigate the hypothesis that CMRV is
fundamentally impossible from a more theoretical standpoint. This
hypothesis will be called ``the impossibility hypothesis'' for short
in the rest of this paper. Before delving into details, we sketch
the main ideas, which come from quantum physics as mentioned in the
Introduction. If the impossibility hypothesis is true, comprehensive
information about the internal working of the cell is not leaking
to the outside environment. This implies that some degrees of freedom
corresponding to the internal workings of the cell are not being observed
in the quantum mechanical sense. This leads to spreading of the quantum
wave packet associated with these degrees of freedom, provided that
these are subject to chaotic time evolution.

To be specific, consider a possibility that the position of a whole
molecule, for example a protein molecule in the cell, is quantum mechanically
smeared to an extent to be discussed. The protein molecule under the
present discussion should not be labeled with a fluorophore because
that would localize the molecule under conditions that enables observations.
In the standard argument, the unlabelled protein molecule is constantly
hit by the surrounding molecules \textemdash{} and hence being ``observed'',
meaning that its quantum state should ``collapse'' to a quasi-position-eigenstate
almost continuously \cite{Tegmark}. However, this argument is focused
solely on the single molecule and all the surrounding molecules are
regarded as environment. In reality, those surrounding molecules are
also quantum objects, which in turn are ``observed'' by yet other
molecules and so on, thus making it difficult to see where the boundary
between the system and the environment is. Naively, such a chain of
``observations'' could lead to accumulation of positional errors.
Also note that, if one molecule is delocalized, then many molecules
must be delocalized in a highly entangled way since the cell is a
crowded place. In the end, however, no comprehensive information about
the positions of these molecules goes outside the cell under the impossibility
hypothesis.

Furthermore, the number of molecules in the cell is basically proportional
to its volume. The amount of information to specify the positions
of all the molecules should be roughly proportional to the volume
if many of those molecules move chaotically and their positions cannot
be deduced. On the other hand, we expect the amount of information
leakage, which allows an outside observer to determine the positions
of molecules, to be basically proportional to the surface area of
the cell. Hence, in this sense, molecules in a larger cell are more
isolated because of the small surface to volume ratio of the cell.
This argument, by itself, leads to an apparently unacceptable suggestion
that larger objects are more quantum, but the strangeness of the argument
does call for a close examination.

Even if we assume that the positions of some molecules are somehow
delocalized in the cell, observable \emph{consequences} of it is unclear.
The thermal de Broglie wavelengths of the molecules are so short that
even supposing that quantum interference was at work, the ``fringe
spacing'' would be extremely small. This raises a point that the
squared wavefunction would most likely be indistinguishable from an
incoherent classical probability distribution. Consequently, \emph{classical},
probabilistic simulation of molecular motion would suffice to model
biological systems, according to this essentially conventional view.
However, we will point out later that quantum mechanical suppression
of otherwise chaotic classical motion of molecules might manifest
itself, despite the short wavelength involved.

In the end, unsurprisingly, these seemingly wild ideas of delocalized
molecules cannot be convincingly supported in the present work. However,
we will find that these ideas cannot be easily dismissed either. That
we cannot conclude one way or another is actually astounding, because
\emph{all} molecules' positions in the cell are well-localized in
space according to the conventional wisdom. In what follows, we will
attempt to make the foregoing argument more precise and, where possible,
semi-quantitative.

\subsection{Preliminary considerations}

To fix thinking, we first consider \emph{Deinococcus radiodurans},
a single-celled life form that withstands the vacuum \cite{Deinococcus radiodurans}.
Suppose that one \emph{Deinococcus radiodurans} cell floats in a vacuum
chamber in a zero-gravity lab on the orbit. We may furthermore suppose
that the vacuum chamber is at the near-zero temperature and the cell
is in the process of being cooled down via radiative cooling, although
it is still at room temperature. This ``setup'', while expensive
practically, simplifies conceptual considerations. Setting aside exotic
possibilities, essentially all information that the cell ``communicates''
to the outside world is sent as infrared (IR) thermal radiation. For
a large enough vacuum chamber we can forget about the evanescent IR
field because it is exponentially suppressed (See Sec. 2.5) and hence
we may consider only traveling IR waves. Abbe's limit then tells us
that a very small amount of information about the workings in the
cell is communicated to us.

One question regarding the above setup is whether the positions of
some molecules in the cell is quantum mechanically well-defined or
not. It is indeed rather likely that the position of the whole cell
is smeared if we recall the experiment observing Young's interference
pattern with the $\mathrm{C}_{60}$ molecule \cite{C60 interference experiment}.
In the experiment, the position of the $\mathrm{C}_{60}$ molecule
is highly delocalized, whereas the relative positions of the constituent
$\mathrm{C}$ atoms are certainly well defined. (The experiment in
Ref.\cite{C60 interference experiment} is well-controlled, in a sense
that the molecular beam was well-collimated and so on, to see the
fringe pattern. In the present situation, we are not talking about
a repeatable experiment to \emph{prove} quantumness of the system.
Hence, here we do not need a well-controlled setup.) Hence, a better
question would be whether \emph{relative} positions of the molecules
in the cell are quantum mechanically smeared. In the case of $\mathrm{C}_{60}$
molecule or a piece of metal, it is safe to say that the relative
positions of atoms are not smeared beyond the conventionally acceptable
values.

Before proceeding, we note that IR radiation from the cell does give
away some information about the state of the cell \emph{thermodynamically}.
This point is easier to see for a piece of cooling metal, rather then
the biological cell, floating in the cryogenic, zero-gravity vacuum
chamber. Measurement of IR radiation allows us to calculate the amount
of reduction of energy $\Delta E$ of the piece of metal due to radiating.
The entropy $S\left(E\right)$ decreases with the decreasing energy
$E$, meaning that the number of available states decreases, thus
increasing our certainty about the state of the system, assuming the
principle of equal \emph{a priori} probabilities of statistical mechanics.
Hence measurement of $\Delta E$ gives us information, although IR
radiation contains basically zero structural information because of
the long wavelength and Abbe's limit. Obviously, this sort of ``general
information'' is not of much interest for us. To see this point better,
next consider a microprocessor chip computing some mathematical function
with power supplied by a small battery, again floating in a vacuum,
enclosed in a tiny metallic case. Measurement of IR radiation emanating
from the metallic box would tell us mostly about the amount of energy
that the battery has lost, \emph{not} the mathematical function that
the processor is computing.

\subsection{Chaotic dynamics as a defining property }

The time evolution of the cell would be predictable if the relative
positions of constituent molecules are fixed, or evolves in a predictable
way. Such a predictable time evolution appears to be at odds with
the impossibility hypothesis because then a finite amount of measurement
to determine the ``initial'' condition would enable CMRV purely
by computation, as in the case of a piece of metal. Hence we are led
to another notion that the defining property of systems like the biological
cell is that the system dynamics is classically chaotic because of
its unpredictability.\footnote{A precise definition of ``classically chaotic'' is not needed here.
We may take it to mean that the classical \emph{simulation} of the
system dynamics using the molecular dynamics (MD), whatever variant
that may be, is chaotic.} Furthermore, it is natural to consider that relative positions of
some molecules becomes quantum mechanically smeared if the system
is classically chaotic under some conditions, as elaborated below. 

An intuitive picture about a classically chaotic system comes from
Hamiltonian classical mechanics, where a volume in the phase space
is preserved; but the volume is exponentially stretched and folded
in a complex way. This means that, in a quasi-classical picture, the
initial wave packet (or more precisely, the Wigner distribution) is
stretched and folded as well, \emph{unless} an external observation
collapses the wavefunction. Dynamical variables, e.g. the relative
positions of molecules in our case, get quantum mechanically smeared
in the absence of such external observations. The phenomenon of quantum
suppression of chaos would set in, when the spread of the wave function
reaches the maximum possible range and start interfering itself. Everything
in this paragraph has been discussed in the literature \cite{Zurek Physica Scripta}.

An immediate counterargument to the above proposition, that the cell
is classically chaotic, may be that the Reynolds' number relevant
to the cell is so small that things do not seem chaotic at all. Although
nothing substantial can be said at the moment besides the argument
invoking the impossibility hypothesis, we point out that the biological
cell is far more complex than a uniform fluid. There are several biological
aspects known to have a very sensitive character. First, the retina
is known to respond to only a few photons \cite{Few photons for retina activation}.
Second, some molecular signaling pathways involve only a few molecules
in the entire cell \cite{Few molecule in signal transduction}. Third,
a chaotic behavior of a nerve cell has been observed \cite{Chaotic behavior of a neuron}.
These aspects may differentiate biological systems from a drop of
water, although properties of water is not fully understood. Evidently,
however, more needs to be studied as to whether biological systems
are classically chaotic or not. In the followings, we will \emph{assume}
such chaotic behavior in order to see where this assumption leads
us to.\footnote{This assumption has been used in a speculative yet interesting argument
on the free will based on unmeasured qubits. See: S. Aaronson, The
ghost in the quantum Turing machine, in \emph{The once and future
Turing: Computing the world}, Ed. by S. B. Cooper and A. Hodges (Cambridge
University Press, 2016). Further study is desirable, especially in
connection with the simulation hypothesis possibly supplemented with
the 5-minute hypothesis. We also note a possibly relevant point that
chaotic quantum systems could be exponentially sensitive to Hamiltonian
parameters. See e.g. M. Combescure, The quantum fidelity for the time-dependent
singular quantum oscillator, J. Math. Phys. $\mathbf{47}$, 032102
(2006); S. M. Roy and S. L. Braunstein, Exponentially enhanced quantum
metrology, Phys. Rev. Lett., $\mathbf{100}$, 220501 (2008).}

We consider the possibility that the cell has chaotic dynamics that
leads to quantum suppression of chaos, which is a quantum phenomenon.
To sharpen our thinking, consider a toy chaotic pendulum with a magnet,
which obviously is totally classical \cite{Chaotic pendulum}. A crucial
question is what differentiates the biological cell from the toy pendulum.
Evidently, the answer is that latter behaves classically because the
pendulum is ``observed by the environment'', such as air molecules
and ambient photons, before its wavefunction spreads. However, one
might press on and ask what happens if the pendulum is enclosed in
a cryogenic vacuum chamber so that an external observer has fundamentally
no chance to extract any information about the pendulum position,
as in the impossible CMRV situation. One might argue that the pendulum
position would be quantum mechanically spread because nobody looks
at it. In fact, it has been argued that even the orientation of Hyperion
the satellite will be quantum mechanically smeared if unobserved \cite{Zurek Physica Scripta}.
In practice, however, minute mechanical friction, or an eddy current
generated by the magnet, \emph{inside the chamber} will cause energy
dissipation that depends on the position of the pendulum, which effectively
constitute an observation in the quantum mechanical sense, thus localizing
the pendulum position.\footnote{While this assertion sounds reasonable, it is difficult to justify
the claim unless we delve into specific details of the setup.} Note that the dissipated energy is stored in degrees of freedom of
e.g. the lattice vibration of the material of the system, which are
not chaotic. Here we consider a possibility, without much support
at the moment, that there are \emph{not} many non-chaotic degrees
of freedom in the cell to dissipate the energy of the chaotic motion
into. Put another way, possibly a large portion of degrees of freedom
of molecules in the cell are themselves chaotic and interacting with
each other, constituting one big and complex chaotic system without
many non-chaotic degrees of freedom left to dissipate energy. Hence,
if quantum suppression of chaos, of the sort that is discussed in
this paper, is to take place, the system under question is \emph{necessarily
microscopic}; as possibly in the case of a molecular machine of life.
This is reminiscent of the contrast between macroscopic rubber balls
in a box (which quickly fall down to the bottom) and air molecules
in a box (which bounce around forever). In this view, the distinction
between the two concepts \emph{macroscopic} and \emph{microscopic}
is not vague and is indeed qualitative in this case --- it amounts
to whether degrees of freedom at smaller scales, to which energy is
dissipated, exist.\footnote{The degrees of freedom at smaller scales considered here should be
\emph{soft}. For example, deformation of nuclei can safely be ignored
because nuclei are hard, or in other words, has a large characteristic
energy scale.} In simple terms, it amounts to whether friction exists or not. Note
that modern integrated circuits do not fall into the category of ``microscopic
machines'' in the sense just mentioned, even though some electronic
circuits are chaotic \cite{Chaotic circuit}.

One might question the above notion that there are not many non-chaotic
degrees of freedoms in the cell to dissipate energy into. About 40-60\%
of cell content is water \cite{Water content in biological systems}
and it \emph{seems} to provide many degrees of non-chaotic freedom.
Although not much can be said about this at the moment, we make a
few remarks for future considerations. First, the entropy $S$ of
water at $25{}^{\circ}\mathrm{C}$ is $70\mathrm{J/mol\cdot K}$ \cite{Water entropy}
and 
\begin{equation}
\frac{S}{N_{A}k_{B}}=\ln W=8.4,
\end{equation}
where $N_{A}$ is Avogadro's number. The number of states $W$ for
a single water molecule is therefore $W=e^{8.4}\cong5\times10^{3}$.
This is a large number, but it should also be kept in mind that a
wavefunction spreads exponentially in chaotic systems, although \emph{only}
in chaotic systems. Second, water exhibits quantum behavior in confined
space \cite{Water quantum effect}, which appears not dissimilar to
the space between crowded molecules in the cell. Third, several interesting
coincidences about the water properties have been pointed out \cite{Sbitnev quantum consciousness}.
Namely, the ``energy broadening'' $\delta E\cong h/\delta t$ due
to the hydronium ion lifetime $\delta t\cong0.2\mathrm{pS}$ in water
is comparable to $k_{B}T=k_{B}\cdot298\mathrm{K}$, suggesting the
quantumness of proton motion. Additionally, the speed of sound is
comparable to the thermal velocity of protons $\sqrt{k_{B}T/m}$ at
room temperature, where $m$ is the proton mass. This suggests that
thermal proton motions easily become collective. While these three
points do not constitute anything solid, they do suggest that our
understanding of water, especially under biological conditions, is
still developing and water cannot easily be regarded as providing
only external, non-chaotic degrees of freedom to make things classical.
In this view, water is \emph{unlike} the solid material of the toy
chaotic pendulum.

A ``quantum measurement'' perspective to see chaotic spread of the
wavefunction followed by interference is the following. The state
of a quantum system collapses to an eigenstate of an observable after
a measurement according to the textbook quantum mechanics. However,
if one sees the system comprising the quantum system and the measurement
apparatus \emph{from the outside}, the whole system, if isolated,
remains in a superposed state even after the measurement. The constituent
states, or in other words the distinct branches of quantum evolution,
normally do not interfere because the measurement record prevents
these from interfering. In fact, in an ordinary macroscopic realm,
the ``measurement record'' get imprinted on e.g. a human brain or
photons flying away from the Earth. \footnote{Tangentially, in some sense the \emph{purification} procedure on a
mixed quantum state may not be a purely mathematical procedure. One
could actually purify a state if he/she manages to physically collect
measurement record somewhere in the world. However, in general that
would be much more difficult than e.g. gathering spilled liquid.} However, there is nothing that prevents interference from happening
if the measurement record is erased by unitary time evolution. In
particular, if the whole system is in a small closed box, then the
number of memory bits in the box may not be sufficient to record the
variety of things that have happened. In this ``saturated'' situation,
distinct events that result in an identical measurement record would
interfere.

However, no system is completely isolated. The state of the system
does collapse if someone makes a measurement from the outside. Consider
a complex chaotic system in a box. Let a complete, orthogonal set
of states deep inside the box be $|a_{1}\rangle,|a_{2}\rangle,\cdots,|a_{n}\rangle$,
where each $|a_{i}\rangle$ represents a classically well-defined
mechanical configuration of parts deep inside the box (or molecular
configurations in the cell). These states would be pointer states
if they were not buried deep inside. Also consider a complete, orthogonal
set of states on and near the surface of the box be $|b_{1}\rangle,|b_{2}\rangle,\cdots,|b_{m}\rangle$,
where each of these likewise represents classically meaningful pointer
states. We assume that one can measure the surface state with respect
to the basis $\left\{ |b_{j}\rangle\right\} $. Let any quantum state
be expressible as a superposition of tensor products $|a_{i}\rangle\otimes|b_{j}\rangle$.
Suppose that the initial state is one of orthogonal states $|a_{k}\rangle\otimes|b_{1}\rangle$,
where the common surface state is taken to be $|b_{1}\rangle$ without
loss of generality. After a certain amount of time duration, each
constituent state $|\psi_{k,initial}\rangle=|a_{k}\rangle\otimes|b_{1}\rangle$
evolves into a complex superposition $|\psi_{k,final}\rangle=\sum_{i.j}c\left(k\right)_{i.j}|a_{i}\rangle\otimes|b_{j}\rangle$
due to chaotic dynamics. Suppose that the measurement on the surface
indicated the state $|b_{p}\rangle$. The measurement does not reveal
which of the states$|\psi_{1,initial}\rangle,|\psi_{2,initial}\rangle,\cdots,|\psi_{n,initial}\rangle$
the system started with, if $\sum_{i}\left|c\left(k\right)_{i.p}\right|^{2}$
depends only weakly on $k$. This last condition seems plausible,
considering the chaotic time evolution. In other words, the ``measurement
records'' going outside are themselves completely jumbled because
of the chaotic dynamics and hence the initial classically meaningful
pointer states $|a_{1}\rangle,|a_{2}\rangle,\cdots,|a_{n}\rangle$
deep inside the box could interfere with each other.\footnote{This is reminiscent of the quantum error correction codes, where the
logical qubit is encoded in physical qubits in a jumbled way so that
measurements on a \emph{part} of physical qubits do not reveal the
content of the logical qubit. Note that this analogy goes only so
far because the feedback part is missing in merely-chaotic quantum
systems.}

We considered mostly the quantum nature of molecular motions. We do
not consider electronic degrees of freedom because the electronic
part only determines the potential energy functions for the positions
of nuclei as far as conditions for the Born-Oppenheimer approximation
are met. Needless to say, this approximation could break down when,
e.g. plasmon-phonon coupling is significant. Although such circumstances
may well play a role in biological systems, considerations of these
are beyond the scope of this paper.

\subsection{The amount of information leakage}

In the above-mentioned case involving \emph{Deinococcus radiodurans},
essentially the only channel through which information about the internal
workings of the cell leaks is thermal radiation. We estimate the information
content of it to gain further theoretical insights about the impossibility
hypothesis. First, we regard thermal photons as independent. The question
is whether internal motion of the cell can be seen via the thermal
radiation through an optical microscope. The answer is obviously no,
but we study this question to gain some insight. We are interested
in molecular details and we take $10\mathrm{nm}$ as the representative
size of a large biological molecule. (For comparison, the prokaryotic
ribosome has the size of about $20\mathrm{nm}$.) We want to see things
at least with this resolution $\lambda_{0}=10\mathrm{nm}$. We consider
the best case, where the emissivity is $1$ at all wavelength. Since
we have $h\nu_{0}\gg k_{B}T$ at this wavelength $\lambda_{0}=c/\nu_{0}$
and at $T=300\mathrm{K}$, Planck's law for the spectral radiance
$B_{\nu}\left(\nu,T\right)$ is approximated as 
\begin{equation}
B_{\nu}\left(\nu,T\right)\cong\frac{2h\nu^{3}}{c^{2}}e^{-h\nu/k_{B}T},\label{eq:plank}
\end{equation}
where $\nu>\nu_{0}$. In the followings precise calculations are not
intended and hence we largely skip calculation of the numerical factor.
We first multiply equation (\ref{eq:plank}) with the solid angle
$\cong4\pi$ into which radiation is emitted, and the specimen surface
area $l^{2}\cong10\left(\mathrm{\mu m}\right)^{2}$. Dividing the
result by the photon energy $h\nu$, we get something that represents
spectral density of photon number flux from the cell 
\begin{equation}
S\left(\nu,T\right)\cong\frac{8\pi\nu^{2}l^{2}}{c^{2}}e^{-h\nu/k_{B}T}.
\end{equation}
We integrate this with respect to $\nu$ from $\nu_{0}=c/\lambda_{0}=c/10\mathrm{nm}$
to infinity to obtain the number of ``high-resolution'' thermal
photons per unit time 
\begin{equation}
\dot{N}\cong8\pi\left(\frac{l}{\lambda_{0}}\right)^{2}\nu_{0}\frac{\alpha\left(\alpha+2\right)+2}{\alpha^{3}}e^{-\alpha}\cong8\pi\left(\frac{l}{\lambda_{0}}\right)^{2}\frac{\nu_{0}}{\alpha}e^{-\alpha},\label{eq:IR photon current}
\end{equation}
where $\alpha=h\nu_{0}/k_{B}T=5\times10^{3}\gg1$ for $T=300\mathrm{K}$
and $\nu_{0}/\alpha=k_{B}T/h=6\mathrm{THz}$ is a characteristic frequency
of the thermal photons. While the pre-exponential factor is $\cong10^{18}\mathrm{Hz}$,
the exponential factor makes $\dot{N}$ effectively zero, so that
we would not get any high-resolution photons even if the photons were
collected for the age of the universe $\sim10^{17}\mathrm{s}$.

Next, we consider another scenario, in which photons are correlated.
This could be the case because the \emph{Deinococcus radiodurans}
cell is isolated in a vacuum and its thermal radiation is not ``random''
but faithfully represents the internal dynamics of the cell that could
involve something similar to ``down conversion'', in which short
wavelength excitations (presumably something akin to phonons) is converted
to IR photons emitted. The resolution beyond Abbe's limit by the factor
$n$, i.e. $\cong\lambda/n$, is in principle possible if, hypothetically,
$n$ entangled photons are involved and a photon detector that detect
$n$-photons together is available \cite{Heisenberg optical resolution}.
Here we suppose most optimistically that\emph{ thermal photons are
entangled in any way we wish}, however unlikely this is, because we
want to estimate a \emph{provable} limit. Since the thermal photon
wavelength at $T=300\mathrm{K}$ is $\lambda_{th}=c/\nu_{th}=5\times10\mathrm{\mu m}$,
we suppose the existence of $n\cong5\times10^{3}(=\alpha)$ photons
in the $n$-particle GHZ state to obtain the resolution of $\sim10\mathrm{nm}$.
We call the group of $n$ photons a ``photon bunch''. For the sake
of simplicity, we suppose all thermal radiation is at the wavelength
$\lambda_{th}$. Stefan-Boltzmann's law suggests that the total power
$P$ from the \emph{Deinococcus radiodurans} cell is 
\begin{equation}
P\cong\sigma l^{2}T^{4}=5\times10^{-8}\mathrm{W},
\end{equation}
where $\sigma$ is the Stefan-Boltzmann constant. Dividing this with
the thermal photon energy $h\nu_{th}=4\times10^{-21}\mathrm{J}$,
where $\nu_{th}=k_{B}T/h$, we obtain the IR photon current $\dot{N}\cong1\times10^{13}\mathrm{Hz}$.
The current is $\dot{N'}\cong2\times10^{9}\mathrm{Hz}$ in terms of
the photon bunch. Hence, $\sim10^{7}$ photon bunches make each image
if we assume a video rate image acquisition of $\sim10^{2}\mathrm{Hz}$.
This number of photon bunches results in a $100\times100$ pixel image
with $\sqrt{10^{3}}\sim30$ dynamic range. This probably is just acceptable
as a CMRV method. However, recall that we made an exceedingly optimistic
assumption that the thermal photons are entangled in a way we want.
It appears quite certain that photons are much less correlated.

Overall, in this rather special case of the \emph{Deinococcus radiodurans}
cell floating in a cryogenic vacuum chamber on a spacecraft, it is
almost certain that information leakage is minimal. If it turns out
that the internal dynamics of the \emph{Deinococcus radiodurans} cell
is not so chaotic, one could replace the cell with a more usual, possibly
more chaotic, cell that is protected from the vacuum with the recently
introduced nano-suit \cite{Nano-suit}. Hence, in view of supporting
the impossibility hypothesis, the present argument based solely on
the IR-radiation is more tenable, though less realistic, than the
rest of the argument in this Sec. 3.3.

We move on to a more realistic scenario, leaving the scenario of the
\emph{Deinococcus radiodurans} cell floating in a cryogenic vacuum
chamber on a spacecraft. The cell is in a standard laboratory in this
realistic case, interacting with the surrounding medium or other cells.
The ``communication channel'' is no longer restricted to IR radiation
and it involves direct ``mechanical'' contacts between the cell
under investigation and the surrounding environment. Considerations
on \emph{phonons} seem more appropriate in this situation. Since the
typical sound velocity is 5 orders of magnitude smaller than the velocity
of light, the parameter $\alpha$ in Eq.(\ref{eq:IR photon current})
is much smaller than $1$ here. Moreover, we expect no significant
acoustic impedance mismatch since everything is rather soft and has
similar density. Hence there could be substantial high-resolution
information going out through the phonon channel.

This suggests that high-resolution biological \emph{phonon} microscopy
could be a promising way forward to attain CMRV \cite{Errico acoustic microscope}.
In fact, acoustic microscopes using superfluid helium have achieved
resolutions at around $10\mathrm{nm}$ \cite{Superfluid helium acoustic microscope}.
Hence the usefulness of the phonon channel in biological imaging under
cryogenic conditions remains to be seen, especially in combination
with RCSP. However, when liquid water is involved as in a living specimen,
the frequency of sound waves with the wavelength $10\mathrm{nm}$
is about $100\mathrm{GHz}$ assuming the sound velocity of $10^{3}\mathrm{m/s}$.
A resolution-limiting factor, among others, in this more typical ultrasonic
imaging is attenuation of the sound at high frequencies, or nonlinearity
at high power used in an attempt to increase the signal to noise ratio
\cite{Ultrasonic microscopy resolution limit}. The standard imaging
theory is useless in such a region of large attenuation and nonlinearity,
where e.g. kinematic approximation is invalid. Hence, we cannot use
arguments or gedankenexperiments that employ the concept of a microscope
to estimate the information leakage.

We need to take a different approach to estimate information leakage
through the supposedly nonlinear and dissipative phonon channel. Since
phonons are about mechanical degrees of freedom, consider the following
gedankenexperiment: To every tiny patch of the surface of the cell
in question, we attach a nano-needle that is a sensitive force sensor
(or a displacement sensor), as shown in Fig. 1 (a). Hence we get \emph{all}
mechanical information emanating from the cell, although whether this
gedankenexperiment can actually be carried out is a separate question
(see Footnote 9). The question is whether we can infer the positions
of $all$ the molecules in the cell from this measurement. Intuitively,
this strikes us as impossible, especially when the size of the cell
is large, because more ``information `` is generated in the cell
than information that can be extracted through the surface because
surface-to-volume ratio tends to be small in a large system.

In the following, we attempt to study if the above intuition is correct.
First, the above argument does not apply to a piece of metal for example,
because of the known atomic structure. However, in a chaotic system
knowledge of the initial condition, even supposing we knew that, would
not allow for knowledge at later times. In other words, the number
of chaotic ``branching'' per unit time\footnote{By chaotic ``branching'', we mean that an initial, unique state
of the system evolves and spreads into two distinguishable states.
This would produce 1bit of uncertainty to the observer.} in the cell is relevant here because we are interested in whether
it is greater than the ``measurement bandwidth'' of the setup in
Fig. 1 (a).

Our situation corresponds to a quantum circuit shown in Fig. 1 (b)
in ``quantum information science'' terms. The quantum circuit represents
a program for a large scale quantum computer that simulates the dynamics
of the cell. The upper portion of qubits, which will be referred to
as \emph{internal qubits}, represent, in a certain encoding scheme,
the degrees of freedom corresponding to the molecules inside the cell,
so that the qubits cannot be measured directly. On the other hand,
the lower portion of qubits, to be referred to as \emph{surface qubits},
collectively represent molecules on the surface of the cell and these
are constantly measured. We assume that the values after the measurement
are not important and hence we simply reset them. The big boxes containing
many quantum logic gates in the circuit represent the chaotic dynamics
under which the cell evolves. 
\begin{figure}
\begin{centering}
\includegraphics[scale=0.25]{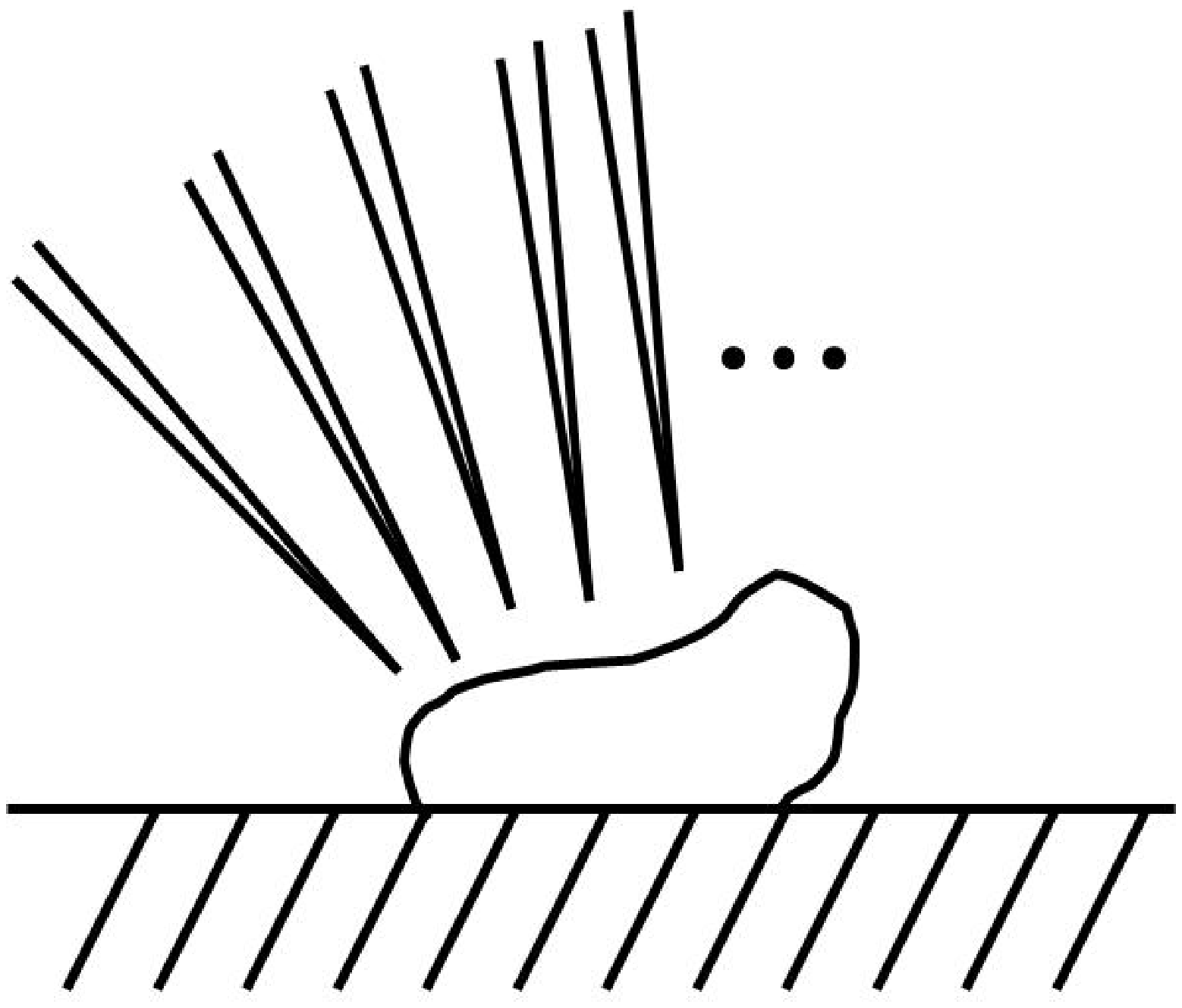}(a) 
\par\end{centering}

\begin{centering}
\includegraphics[scale=0.5]{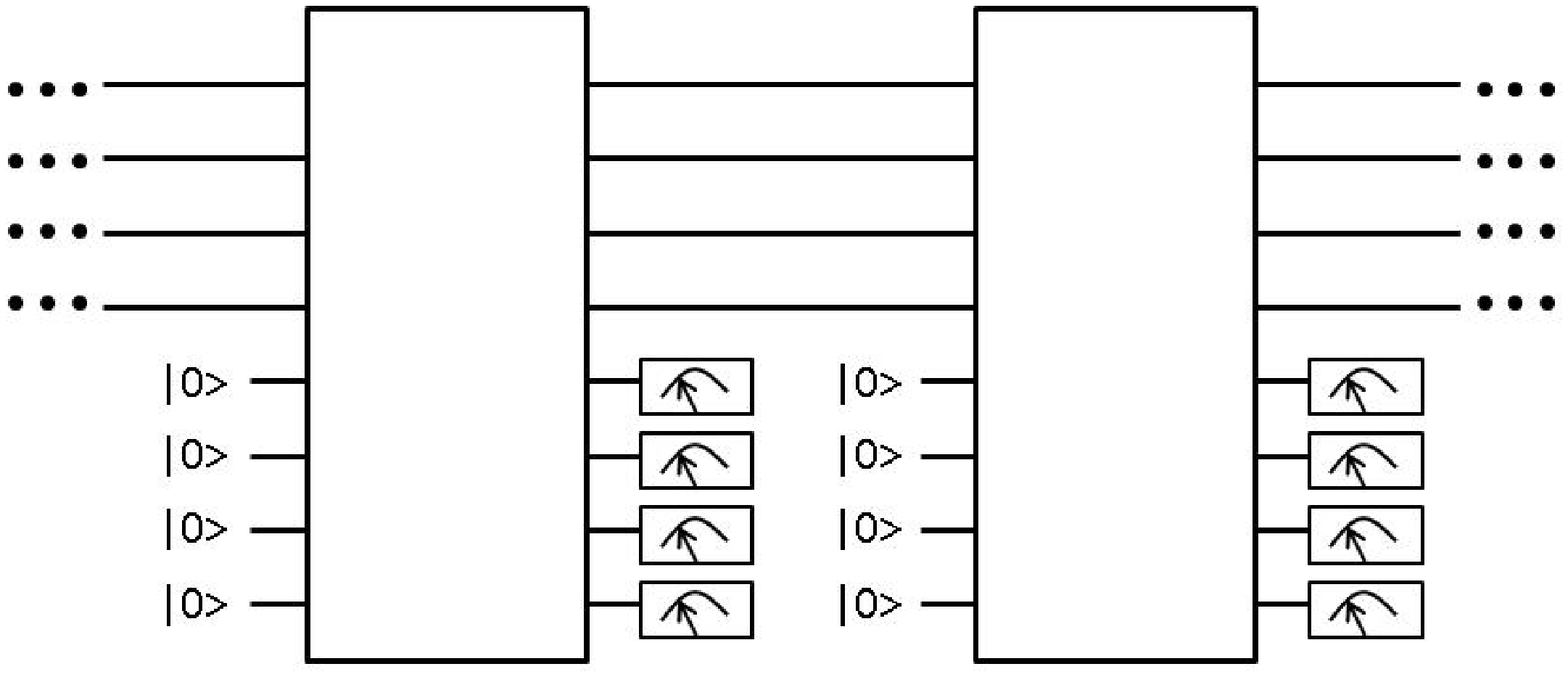}(b) 
\par\end{centering}

\caption{(a) Comprehensive mechanical measurement on a biological cell in our
gedankenexperiment. A number of force/displacement sensors (the needle-like
objects) are placed on the cell (the smooth object on the substrate).
The substrate may have to incorporate many sensors as well. (b) A
quantum circuit model of the cell. The lower portion of the qubits,
corresponding to the degrees of freedom on the cell surface, are constantly
measured (measurements are represented by the meter objects) and reset
to the state $|0\rangle$. The upper portion, corresponding to the
internal degrees of freedom of the cell, is only indirectly accessible
from the measurement apparatus, i.e. from the surrounding environment.
The boxes contains quantum circuit describing the chaotic time evolution.
The number of qubits is arbitrary, but here we show 8 qubits for clarity.}
\end{figure}

The internal dynamics of the cell --- or in terms of the quantum circuit
shown in Fig. 1 (b), the circuit structure inside the large box ---
has to be considered to estimate the amount of information leakage.
The reason is simple: For an extreme example, the state of the internal
qubits is certainly unobservable if no quantum logic gate connects
the internal qubits and the surface qubits. On the other hand, consider
an internal qubit A corresponding to a degree of freedom pertaining
to a molecule deep inside the cell, which interacts indirectly with
a surface qubit Z through a chain of qubits, which roughly represents
a chain of molecules that has something akin to nearest-neighbor interactions.
Suppose for simplicity that in this chain of qubits $\left\{ \mathrm{A,B,C,\cdots Z}\right\} $,
the interaction is such that qubit A controls qubit B with the controlled-NOT
(CNOT) gate, and then qubit B controls qubit C, and so on, eventually
reaching the surface qubit Z. Further suppose that initially the qubit
A is in the symmetrically superposed state
\begin{equation}
|\psi_{A}\rangle=\frac{|0\rangle+|1\rangle}{\sqrt{2}},\label{eq:symmetric state}
\end{equation}
where $|0\rangle$ and $|1\rangle$ correspond to two localized states
of the internal molecule, and the rest of the qubits are in the state
$|0\rangle$. Then we eventually get an $n$-qubit GHZ state, on which
if a measurement on the surface qubit Z is made with respect to the
basis $\left\{ |0\rangle,|1\rangle\right\} $, then the entire wavefunction
collapses. Hence this is the opposite extreme, in which the internal
information of the cell is obtainable by measuring the surface molecules.
Note that this second situation is at odds with the impossibility
hypothesis, except that this particular quantum circuit may not allow
for \emph{comprehensive} internal information acquisition because
of the existence of other qubits etc. There are of course cases in
between. For example, if each of the above CNOT gate is replaced with
a similar gate that rotates the controlled qubit only with a small
angle $\varepsilon$ as opposed to the angle $\pi$, the probability
amplitude for the state in which the qubit Z is in $|1\rangle$ is
approximately $\varepsilon^{N-1}$, where $N$ is the number of qubit
in the chain. One may regard this as exponential suppression of information
leakage that is fully consistent with the impossibility hypothesis.

Thus, these considerations suggest that detailed knowledge of the
biological machinery in the cell is needed to estimate the information
leakage rate. However, one may hope go beyond this statement to gain
at least a rough estimation without delving into biological details.
In the followings we make several crude attempts to this end.

First, we show that the above-mentioned exponential suppression is
quite unlikely to happen in the cell for the following reasons. Consider
two molecules colliding each other in the cell. Assume for simplicity
that these molecules are effectively rigid. If the collision is weak
enough to enable the exponential suppression, it should leave the
final quantum state of the two molecules largely overlapped with the
initial state. This may be seen most clearly by the ``caricatured''
(in the sense of the word used by Zurek \cite{Zurek Physica Scripta})
version appeared above, where the angle-$\varepsilon$ CNOT gate illustrates
the weak molecular collision. On the other hand, an actual molecular
collision should typically transfer energy of the order of $k_{B}T$
simply because the molecules have such a range of varying energy.
However, there are many ``energy levels'' within the width $k_{B}T$,
resulting in a nearly complete change of the quantum state upon collision.
(Note that the positional degrees of freedom of the molecule have
continuous energy levels. However, they are \emph{effectively} discretized
to the ``energy levels'' with a spacing $\sim\Delta E$, where the
value of $\Delta E$ corresponds to the positional localization length
of the molecule.) This is how a version of $\hbar\omega\ll k_{B}T$
comes up here, which is desirable because any discussion lacking emergence
of this aspect would be missing an important aspect of large thermal
energy. Nevertheless, this shows that the idea of exponential suppression
occurring in the cell is most likely to be incorrect.

Since weak collisions between the molecules are unlikely, we need
other models of the cell that involves ``strong'' collisions. First,
consider a model, in which neighboring molecules exchange their quantum
states upon collision, much like the SWAP gate in a quantum circuit.
Hence in this model, a piece of quantum information random walks and
hence the information will have leaked when it hits the surface of
the cell. When considering this, we imagine localized molecules colliding
with each other exchanging their states, which is at odds with another
mental picture, in which the molecular positions are delocalized to
cause quantum suppression of chaos. Let us ignore this apparent inconsistency
and plow ahead. Observe that this whole process is like a random walk,
in which changes of the direction are caused by intermolecular collisions.
Hence, the ``diffusion constant'' associated with the process is
probably not too different from the actual diffusion constant of biological
molecules in a cellular environment. Past measurements \cite{Diffusion constant of biomolecules}
have shown that such diffusion constants are roughly in the range
of 
\begin{equation}
D\cong10^{-6}\mathrm{cm^{2}/s}=\left(1\mu\mathrm{m}\right)^{2}/\left(10\mathrm{ms}\right).
\end{equation}

Another model we consider employs a ``random unitary operator''
as a single-step discrete time evolution operator. In other words,
each big white box in Fig. 1 (b) is such a random unitary operator.
We mention a few caveats before we start. First, our definition of
the ``random unitary operator'' $R$ is that the absolute values
of all the upper triangular matrix elements and the diagonal matrix
elements are independent and identically distributed (i.i.d.) random
numbers, while the phases of all upper triangular matrix elements
are drawn from the uniform distribution in $[0,2\pi)$. In addition,
$R$ of course obeys the unitarity relation $R=R^{\dagger}$. It must
be noted that the usual definition of ``random unitary operator''
is much more sophisticated in order to make certain properties invariant
under relevant transformations \cite{Dyson random matrix,Random unitary matrix generation}.
However, our discussion is intended only to make a first dent into
the problem and hence we do not use advanced concepts. See, e.g. Ref.\cite{Quantum Supremacy}
for an example of more advanced treatment. Second, there are ``exponentially
many'' states in the Hilbert space (although technically the number
is uncountable infinity) and hence there are correspondingly many
unitary operators that bring $|0\rangle$ to these states. Evidently,
most of these unitary operators contain exponentially many quantum
logic gates. Therefore, the actual unitary operator that govern cellular
time evolution should not be randomly drawn from all the possible
unitary operators, but from a ``polynomially large'' set.\footnote{Here we abuse the terminology from computational complexity theory
in order to present a rough idea.} We will largely ignore this issue. Third, since the operator is random,
our argument ignores geometric structures of the cell. It does not,
for instance, take the 3-dimensional nature of the space into account.
A randomly distributed nearest-neighbor CNOT gates in a 3-dimensional
lattice, for example, would constitute a better model but that is
beyond the scope of the present work.

Having mentioned these caveats, we proceed to use the quantum circuit
model incorporating the random unitary operator. Note that the general
quantum circuit model can describe \emph{any} quantum system, up to
mathematical technicalities. To be specific, let the number of all
qubits and the internal qubits in Fig. 1 (b) be $n$ and $m$, respectively.
Hence the number of surface qubits is $n-m$. Let $N$ be $N=2^{n}$
and likewise $M=2^{m}$. Basis states are written as $|b_{1}b_{2}\cdots b_{n}\rangle=\bigotimes_{i=1}^{n}|b_{i}\rangle_{i}$,
where $b_{i}$, whose value is either $0$ or $1$, represents the
value of $i$-th qubit, whose state space is spanned by $|0\rangle_{i}$
and $|1\rangle_{i}$. The label $i$ starts from $1$ to $n$ and
points to a surface qubit when $n-m<i$. An alternative way to represent
a basis state is $|k\rangle$, where binary representation of the
integer $k$ is $b_{1}b_{2}\cdots b_{n}$, so that $0\leq k\leq N-1$.
We normalize the state vectors so that $\langle k|l\rangle=\delta_{k,l}$.
The complex coefficient of the basis state $|k\rangle$ is written
as $c_{k}$. The random unitary operator $R$ representing each big
white box in Fig. 1 (b) has matrix elements $R_{k,l}=\langle k|R|l\rangle$.
We formulate the problem as follows. Let the first qubit $|b_{1}\rangle_{1}$
represent the internal dynamics of interest, so that $|0\rangle_{1}$
and $|1\rangle_{1}$ represents the position of the molecule under
consideration. Suppose that the position of the molecule is already
spread due to the chaotic dynamics and hence the state is $\left(|0\rangle_{1}+|1\rangle_{1}\right)/\sqrt{2}$.
Furthermore, without loss of generality, we set all other qubit in
the state $|0\rangle_{i}$, where $i=2,3,\cdots,n$. To avoid excessive
complexity in our analysis, assume that all qubits are disentangled
at the beginning. Let the encoding, or mapping, of the molecular quantum
states into the qubit states be such that the state $\bigotimes_{i=2}^{n}|0\rangle_{i}$
translates to having a molecular quantum state that is in accordance
with the conventional wisdom --- that all molecules except the one
described by $\left(|0\rangle_{1}+|1\rangle_{1}\right)/\sqrt{2}$
are localized classically and these are not quantum mechanically entangled.
Hence, the quantum state we initially have is
\begin{equation}
|\varPsi\rangle=c_{0}|00\cdots0\rangle+c_{1}|10\cdots0\rangle,\label{eq:superposed initial state}
\end{equation}
where $c_{0}=c_{1}=1/\sqrt{2}$. This is a superposition of the positional
states of the molecule being at two different places. Here we ignore
the apparent inconsistency that only one molecule is delocalized in
the crowded cellular environment. The question is how well this superposition
of equal amplitude is preserved after a single step of time evolution
and a measurement on the surface qubits. We apply the unit-time-step
evolution operator $R$ to Eq.(\ref{eq:superposed initial state})
to obtain
\begin{equation}
\frac{R_{0,0}|0\rangle+R_{1,0}|1\rangle+R_{2,0}|2\rangle+\cdots}{\sqrt{2}}+\frac{R_{0,1}|0\rangle+R_{1,1}|1\rangle+R_{2,1}|2\rangle+\cdots}{\sqrt{2}},\label{eq:superposed final state}
\end{equation}
where the first term represents the state evolved from the first term
in Eq.(\ref{eq:superposed initial state}) and likewise for the second
term. A measurement on all $n-m$ surface qubits results in only $M$
remaining terms in the numerator of each term. Up to a normalization
factor, we have
\[
|A\rangle+|B\rangle=\left\{ R_{X,0}|X\rangle+R_{X+1,0}|X+1\rangle+R_{X+2,0}|X+2\rangle+\cdots\right\} 
\]
\begin{equation}
+\left\{ R_{X,1}|X\rangle+R_{X+1,1}|X+1\rangle+R_{X+2,1}|X+2\rangle+\cdots\right\} ,\label{eq:state after measurement}
\end{equation}
where $X$ depends on the measurement outcome and the states $|A\rangle,|B\rangle$
represent respectively the right-hand-side terms in the first and
second set of braces. There are several points to consider. The first
point is about the orthogonality between the states $|A\rangle$ and
$|B\rangle$. Since the matrix $R_{k,l}$ is ``random'', we first
make an overall normalization to the matrix so that the average of
the square of the absolute value is $\left\langle |R_{k,l}|^{2}\right\rangle \approx1/M$.
Hence $|A\rangle$ and $|B\rangle$ have approximately the unit length.
We obtain $|\langle A|B\rangle|\cong1/\sqrt{M}$ unless $R_{k,l}$
obeys an unusual probability distribution, because the summation goes
in the ``random walk'' fashion due to the random phase of $R_{k,l}$.
Hence, $|A\rangle$ and $|B\rangle$ are approximately orthogonal.
The second point is how well the equality of the probability amplitudes
of $|A\rangle$ and $|B\rangle$ is preserved. This depends on the
probability distribution, to which $R_{k,l}$ obeys. For example,
if $|R_{k,l}|=1/\sqrt{M}$ holds exactly and $R_{k,l}$ only have
random phases, the equality of the amplitudes of $|A\rangle$ and
$|B\rangle$ is preserved perfectly. In general, the equality of the
probability amplitudes is preserved approximately, again unless $R_{k,l}$
obeys an unusual probability distribution. The third point is whether
the relative phase between $|A\rangle$ and $|B\rangle$ is preserved.
As can be verified easily, the relative phase is not preserved at
all. It is unclear if such an introduction of random phase affects
the quantum suppression of chaos. This circumstance resembles to,
if we take an analogy from condensed matter physics, e.g. a coherent
electron wave under a fluctuating external magnetic field, as opposed
to a dissipative motion of the electron that gets entangled with the
lattice degrees of freedom. The reason is that the states $|A\rangle$
and $|B\rangle$ remain largely coherent and do not evolve into a
mixed state despite the phase noise.

Although the above analysis seems to support the notion that the information
leakage is small, this result should be taken with a grain of salt.
First, the random unitary operator which brings all states to all
states in one time step is not at all realistic. The real-world time-evolution
is likely to be sparse with respect to a suitable basis. Second, the
above results, such as the difference in probability amplitudes between
$|A\rangle$ and $|B\rangle$, depend only on $M$ and not on $N$.
This unacceptable aspect comes probably from the fact that we used
a rudimentary definition of ``random unitary matrix''. Moreover,
our analysis does not give us an actual time scale. It \emph{appears}
natural to take roughly the time scale of an inverse Lyapunov exponent
to carry out the single time evolution by the operator $R$. However,
we do not know the values of such relevant Lyapunov exponents, or
how many of them exists, let alone the question of whether they exist
at all.

\subsection{Quantum suppression of chaos}

In view of the above difficulty in estimating the information leakage
rate, it is even more difficult to estimate how the suppression of
chaos in the cell might play out. Hence we will be brief. We continue
to assume that the dynamics of the cell is classically chaotic and
that the information leakage rate determines how the quantum state
of the cell collapses. The process is, in a sense, competition between
the chaotic expansion of the quantum state in the positional space,
which occurs with the rate proportional to characteristic Lyapunov
exponents, and the collapse of the quantum state by external observations
that occurs with the rate that is essentially proportional to the
information leakage rate. As mentioned above, we do not know how many
unstable degrees of freedom exist in the classical model of a biological
cell, nor the various positive Lyapunov exponents associated to these
degrees of freedom. A preliminary study could start once these are
more or less known. Then we might model the process of ``external
observation'' by collapsing one randomly-selected degree of freedom,
for example, setting the ``spread of the wave function'' (SWF) associated
with the degree of freedom to the initial value. When the SWF reaches
its maximum value, which will be referred to as the ``range'' below,
we may regard that as beginning of quantum suppression of chaos \cite{Zurek Physica Scripta}.

To see the crudest physics, consider a simplified system, in which
all the positive Lyapunov exponents $\lambda_{i}$ and the range $L_{i}$
of the associated dynamical variables $x_{i}$ are the same for all
$N$ unstable degrees of freedom. Hence $\lambda_{i}=\lambda$ and
$L_{i}=L$ for all $i=1,2,\cdots,N$. Simplifying even further, the
equations of motion that we consider is a set of \emph{independent}
differential equations 
\begin{equation}
\dot{y}_{i}=\lambda y_{i},
\end{equation}
where $y_{i}$ is roughly the SWF associated with the $i$-th degree
of freedom. (For example, the $x_{i}$-dependence of the wavefunction
at some given time may approximately be $\simeq e^{-x_{i}^{2}/2y_{i}^{2}}$.)
Hence we have $y_{i}=y_{0}e^{\lambda t}$, where $t$ measures the
elapsed time since the last wavefunction collapse and $y_{0}$ is
the characteristic size of the wavefunction right after the collapse.
Note that the wavefunction spreads fully in the allowed range $L$
of the variable $x_{i}$ at $t=t_{L}\cong\left(1/\lambda\right)\ln\left(L/y_{0}\right)$
and the suppression of chaos begins. Note the insensitivity of $t_{L}$
on $L$ and $y_{0}$. Suppose that each degree of freedom is ``measured''
with a time interval $\Delta\tau$. Then we see that it is a sort
of ``speed competition'' between the exponentially expanding SWF
and its collapse caused by the external observations. We might conjecture
that this speed competition results in two phases in general, such
that the chaotic expansion dominates in one phase ($t_{L}\ll\Delta\tau$)
while measurement-induced collapses dominate in the other ($t_{L}\gg\Delta\tau$),
depending on the rate of the external observation $\Delta\tau^{-1}$.

\subsection{Possible places to find experimental evidence}

It is rather easy to \emph{imagine} a highly conclusive experimental
proof for large-scale quantumness in a biological system.\footnote{For example, it would be a quite conclusive proof if a man factors
large integers super-fast. Less extravagantly, a detection of a violation
of any version of Bell's inequality by measuring quantities at two
separate places on a biological system will also constitute a strong
proof. It would indeed be conclusive if the separation of two measurement
events is spacelike. However, even an impossibility of signaling by
the speed of some biologically relevant phenomena (nerve signaling,
speed of sound, ``speed'' of molecular diffusion etc.), as opposed
to the speed of light, in an experiment would constitute a rather
convincing demonstration.} However, at present we are not able to propose an experimental scheme
that has a non-zero chance to generate evidence for/against quantum
suppression of chaos in a biological cell with a reasonable degree
of certainty. Hence, we content ourselves with considering where we
might find at least circumstantial evidence.

First, there may be thermodynamic evidence. Recall that von Neumann
entropy does not increase in an isolated quantum system, even in a
system apparently out of equilibrium \cite{Quantum info and thermodynamics}.
This apparent paradox is resolved when interaction with the environment
is taken into account. In a usual system, the positional degrees of
freedom is subject to einselection, i.e, environment-induced superselection.
Hence, the exponential spread of the state in a chaotic system is
reflected as a linear increase of entropy as far as the system is
observed frequently enough. This process is obviously affected if
quantum suppression of chaos is at work in the cell. We check some
numbers: Human body comprising $\sim10^{13}$ cells \cite{Human body cell count}
produces heat flux on the order of $100\mathrm{W}$, suggesting that
each cell produces $\dot{Q}\cong10\mathrm{pW}$ of heat flux. From
the relationship 
\begin{equation}
dQ=TdS=k_{B}T\cdot d\left(\ln W\right)=0.69k_{B}T\cdot d\left(\log_{2}W\right),
\end{equation}
where $S$ is entropy, we can compute how many bits of uncertainty
are produced in a cell per unit time as
\begin{equation}
\frac{d\left(\log_{2}W\right)}{dt}=\frac{\dot{Q}}{0.69k_{B}T}\cong10^{9}\mathrm{bit/s},
\end{equation}
where $T=300\mathrm{K}$. If the size of a cell is roughly $\left(1\mu\mathrm{m}\right)^{3}$,
then the above calculation suggests that 1 bit of uncertainty, or
chaotic spread by a factor $2$ in our picture, is generated in a
volume $\left(1\mathrm{nm}\right)^{3}$ every second. For instance,
if this rate turns out to be too small for the supposedly chaotic
molecular motions in the cell, this number would support the hypothesis
that quantum suppression of chaos is going on in the cell. However,
all this depends on how many chaotic degrees of freedom there are
and what Lyapunov exponents are associated to these. Hence further
investigations into the classical dynamics in the cell are desired.

The reader might feel that the concept of quantum suppression of chaos
is too eagerly invoked in the above, when heat is simply generated
from another form of energy. As a sanity check, consider a light bulb
and see how many bits of uncertainty is produced per unit time, similar
to the above calculation. We indeed obtain, from the value of heat
flux $\dot{Q}$, the number of bits that a single tungsten atom contributes:
\begin{equation}
\frac{d\left(\log_{2}W\right)}{dt}=\frac{\dot{Q}}{0.69k_{B}T}=\frac{\rho\Delta V}{0.69k_{B}T}\left(\frac{I}{\pi d^{2}/4}\right)^{2}\cong10^{2}\mathrm{bit/s},
\end{equation}
(where $T=10^{3}\mathrm{K}$, the tungsten wire diameter is $d=50\mathrm{\mu m}$,
the resistivity is $\rho=25\mathrm{\mu\Omega\cdot cm}$, the volume-per-atom
is $\Delta V=1.6\times10^{-29}\mathrm{m^{3}}$ and the current is
$I=1\mathrm{A}$.) This number has nothing to do with the chaotic
spread of a wavefunction. Instead, it reflects the following situation:
The ``motion'' of a tungsten atom is excited by the electric current,
thus having more quantum states available. Then the atom releases
heat to the environment, with which ``uncertainty'' is thrown away
because a smaller number of quantum states are available for the tungsten
atom after releasing heat. In the case of chaotic systems, the entropy
increase due to the spread of wavefunction is understood only in terms
of interaction with the environment, that converts the initially pure
state (for the sake of argument, \emph{imagine} that we measured the
state of the system before the wavefunction spread) to a mixed state.
Hence it is natural to consider that the measurement process itself
injects the necessary energy to the chaotic degree of freedom. In
the biological cell under our hypothesis, this energy comes presumably
from internal degrees of freedom that ``monitors'' the chaotic degree
of freedom. Eventually, the heat must be released to the environment
surrounding the cell. Also recall that, at the same time, we put forward
the notion that there are not many degrees of freedom that monitor
the putative chaotic degrees of freedom. Hence, under this notion,
the time for the wavefunction spread is sufficiently long that quantum
suppression of chaos can set in. In other words, the ``monitoring''
mentioned above occurs infrequently enough so that the heat generation
is suppressed.

Second, if functioning of the cell, or functioning of life more broadly,
critically depends on quantum suppression of chaos, then classical
simulations, e.g. MD simulation, of biological cell should fail. A
difficulty with this ``approach'' is that there should be myriad
of other things that could make simulations fail. Conversely, falsification
of the existence of quantumness in the cell, providing that is the
case, would be relatively straightforward by successful classical
simulations.\footnote{Thus, falsification is easier than a proof in this particular case
because one needs only a single example of successful classical simulation
to do so. However, conceivably we could adapt a method from the ``quantum
supremacy'' research \cite{Quantum Supremacy} to give a single example
of successful quantum \emph{experiment} on an actual biological system.} On a related note, RCSP described in Sec. 2.1 ``classicalizes''
the system at each time step separated by $\Delta t$ and this is
in a sense similar to classical simulation. Hence, RCSP should also
fail to reproduce biologically informative time evolution under the
hypothesis that quantum suppression of chaos is at work in the cell.

\section{Conclusions and afterthoughts}

We started our study by evaluating the impossibility hypothesis, which
asserts the fundamental impossibility of a certain type of comprehensive
imaging (i.e. CMRV) of the living cell. We were then led to the unusual
notion of the possible existence of quantum mechanically delocalized
molecules in the biological cell because of the lack of quantum mechanical
monitoring, which in turn brought us to the notion of quantum suppression
of the putative classical chaos in the biological cell. Our study
did not find solid evidence supporting these ideas. Our rather weak
conclusion only says that \emph{we currently do not know }if such
a quantum effect exists. Nonetheless, we obtained the following two
findings. First, we found that the question about the existence of
delocalized molecules in the cell is worth asking at all. In fact,
this question has hardly been asked before. Second, we found that
answering this question is hard. This is a non-trivial finding because
the answer is \emph{obvious} according to the conventional wisdom,
which says \emph{all} molecules are localized in the cell.

The present study is built on quantum chaology research, in which
it is known that even huge objects can evolve into a quantum mechanically
smeared state if the system is classically chaotic, and if the system
is not observed (i.e. without decoherence). We argued that experience
shows that it is difficult, or perhaps impossible, to measure the
internal states of the cell comprehensively, real time, at molecular
resolution and in true 3D. This suggests that external observers cannot
observe some internal states of the cell. We argued that in the cell,
unlike other chaotic systems that have been studied, it could be that
there are not many non-chaotic degrees of freedom that acts effectively
as an internal observer. The reason is that the cell could be a \emph{microscopic}
molecular machine without many extraneous degrees of freedom. We argued
that, if that is the case, the chaotic, exponential spreading of wavefunction
may outpace the rate at which decoherence effects ``collapse'' the
wavefunction. We argued that, in spite of the short wevelength of
the matter wave and the physiological temperature involved, quantum
effects, if they exist, could manifest itself through the known phenomenon
of quantum suppression of otherwise classically chaotic dynamics.
We showed that, if the phenomenon is actually there, it might leave
its fingerprints in some places although none of the suggested ``fingerprints''
are conclusive.

We reiterate the above summary, this time from another viewpoint of
how the biological cell, a warm-and-wet physical system, could evade
various mechanisms of the emergence of classicality. The first mechanism
of the emergence of classicality is similar to how ray optics emerges
from wave optics, making the position of an object well-defined. In
our case, however, a quantum wavepacket could spread fast enough because
of the putative chaotic dynamics. The spread of wavepacket in turn
causes interference eventually, despite the fact that the de Broglie
wavelengths of molecules are far shorter than the length scale associated
with the potential. The second mechanism works when energy associated
with the system, such as $k_{B}T$, is sufficiently large to obscure
individual energy quanta $\hbar\omega$. However, it is known in quantum
chaology that quantum effects manifest itself even when the energy
of an object is very large and the associated de Broglie wavelength
is short, \emph{if} the object is well-isolated. This brings us to
the third mechanism of decoherence. We argued that if there are a
sufficient number of chaotic degrees of freedom in the volume of the
cell, information about the motion of these degrees of freedom may
not leak out effectively because of the limited surface area of the
cell. This could make the degrees of freedom somewhat isolated, which
is a prerequisite for the suppression of decoherence.

Before moving on to considerations on possible consequences of the
quantumness we have been discussing, we consider a couple of alternative
scenarios. First, it should be pointed out that chaotic dynamics is
not the only way to amplify quantum uncertainty, or to spread the
wavepacket. As exemplified by the Schrodinger's cat argument, sensitive
measurement apparatus can play a similar role. Dynamics of the cell
could simulate chaotic dynamics if some molecular systems in the cell
effectively act as sensitive quantum measurement apparatus. Note that,
however, technically such a system would not necessarily be chaotic
in the strict sense because chaoticity requires exponential divergence
of classical trajectories at essentially everywhere. Nonetheless,
such strict chaoticity is not required in the argument presented in
this paper. Second, it could be that CMRV is impossible \emph{not}
because there is no information leaking out. Instead, it might be
that the leaked information is hopelessly jumbled up at the cell surface
and ``reconstruction'' of the desired information is next to impossible.
If this is the case, quantum coherence needed for the quantum suppression
of chaos does not exist. At the same time, this possibility would
raise the hope that CMRV is possible after all, perhaps using massive
computing power. However, it should be mentioned that this scenario
appears to be at odds with the ``scaling'' idea that the surface
area may not be large enough to transmit information arising from
the number of ``chaotic information sources'' that is presumably
proportional to the volume, as the system size increases. Nonetheless,
it may be that chaotic information sources are not independent but
entangled to each other, so that their number is not proportional
to the volume.

We next consider possible consequences of the truth or falsehood of
the hypothesis that molecules are delocalized in the cell. If it turns
out that all molecules in the cell are well-localized as the conventional
wisdom states, then there will be no change in our conception of biological
systems. Things may have to be reconsidered, however, if it turns
out that some molecules are delocalized and quantum suppression of
chaos is at work. For example, future research programs of simulating
the whole cell by MD simulation will have to take this effect into
account, or at least keep the effect in mind. It is currently unknown
whether the quantum-mechanically suppressed chaotic systems can be
efficiently simulated with a classical computer. If it is impossible
to classically simulate such systems efficiently, that might represent
an obstacle in the field of computational biomodeling.\footnote{Explicitly showing the impossibility to efficiently simulate a quantum
chaotic system with a classical computer amounts to proving $\mathrm{BPP}\neq\mathrm{BQP}$,
which is regarded difficult.} On the ``brighter'' side, such impossibility will make quantum
computing research more relevant to life science than it already is.
More speculatively, future artificial ``synthetic biology'' systems
might perform otherwise intractable computations in this case.

Suppose that the notion appeared in our study --- the classically
chaotic dynamics of molecules in the cell is suppressed quantum mechanically
--- is actually true. An interesting question is whether this has
\emph{direct} biological implications, as opposed to implications
regarding biology \emph{research}. In the field of physics of computation,
it has been argued that all computations can be done without spending
energy, by way of \emph{reversible computing} \cite{Nielsen Chuang}.
A theoretical example of such a reversible computer is the ``billiard
ball computer'', in which frictionless motion of colliding classical
balls accomplishes any computation without spending energy. There
is a caveat, however: The billiard ball computer is extremely sensitive
to an external force, internal friction and so on; and its dynamics
is essentially chaotic. To make it work, one has to set up feedback
systems to \emph{measure} the positions of the balls, and if a ball
goes away from the supposed trajectory, the feedback system should
gently push the ball back on the track. However, measurement entails
measured data stored in a memory, which eventually need to be discarded.
Then, the Landauer-Sagawa-Ueda theorem \cite{Landauer's principle,Sagawa-Ueda}
states that heat must be generated in this whole process of measurement
and memory erasure, or in other words entropy must be thrown away.
One question associated with this is whether the chaotic motion of
a reversible computer can be suppressed quantum mechanically at least
partially, thus reducing the reliance on feedback systems. \emph{If}
that is possible, a fascinating question then is whether such a mechanism
is at work in biological systems to maintain their ``orderly dynamics''
without generating excessive heat.\footnote{The ongoing advances in artificial intelligence might offer a clue
that, however, is admittedly tenuous. For instance, the 2016 match
of the game Go between a top Go professional Lee Sedol and an artificial
intelligence (AI) AlphaGo, while Sedol lost, showed that a human professional
and a computer had at least comparable performances at the time. Here
we attempt to compare the amount of energy taken for both sides for
\emph{training}, as opposed to the game playing itself (in which AI
consumes much larger power). The human brain consumes about $20\mathrm{W}$
of power, and a popular theory posits that it takes $10^{4}\mathrm{hours}\sim4\times10^{7}\mathrm{s}$
of time to become an expert (M. Gladwell, \emph{Outliers}, Little,
Brown and Company 2008), resulting in training energy of roughly $10^{9}\mathrm{J}$
for a human being. On the other hand, AlphaGo uses $\sim50$ processors,
each of which consume perhaps about $\sim30\mathrm{W}$ of power,
and was trained by self-playing for a month ($\sim3\times10^{6}\mathrm{s}$).
(Note also that a ``distributed'' version comprises $\sim10^{3}$
processors. See the reference at the end of this footnote.) This indicates
training energy of about $5\times10^{9}\mathrm{J}$ for an AI. These
two values are comparable and we may at least state that difference
in energy efficiency of many orders of magnitude does not exist between
these biological and solid state computing systems. Nonetheless, in
view of the crudeness of this estimation (we did not even speculate
about the likely algorithmic difference between them), further study
is desirable. See: C. S. Lee et al. Human vs. computer Go: review
and prospect, IEEE Comput. Intell. M., $\mathbf{11}$, 67 (2016);
D. Silver et al. Mastering the game of Go with deep neural networks
and tree search, Nature $\mathbf{529}$, 484 (2016).}

Apart from these speculations, even if the idea studied in this work
--- the classically chaotic dynamics of molecules in the cell is suppressed
quantum mechanically --- turns out to be wrong, which definitely is
possible, the following idea should be independent from that: There
can be a classically chaotic system, which is microscopic in the sense
we discussed, whose internal dynamics is highly unobservable from
the outside, whose chaotic motion is quantum mechanically suppressed
before decoherence sets in, and which operates quantum mechanically
outside the traditional domain of quantum systems. Such an engineered
system, if built, could find useful applications that might include
ultralow-power computing.

\section*{ACKNOWLEDGMENTS}

This research was supported in part by the JSPS ``Kakenhi'' Grant
(Grant No. 25390083).

\end{document}